\pdfoutput=1

\documentclass[table,xcdraw,sigconf,authorversion,nonacm,balance=false]{acmart}

\usepackage[nolist]{acronym}
\usepackage{amsmath}
\usepackage{amsfonts}
\usepackage{csquotes}
\usepackage{enumitem}
\usepackage{multicol}
\usepackage{tikz}
\usepackage{threeparttable}
\usepackage{xspace}

\setlist[itemize]{leftmargin=*}
\setlist[enumerate]{leftmargin=*}

\usepackage[capitalise, nameinlink, noabbrev]{cleveref}

    \newcommand*{\commit}               {\textsf{Commit}\xspace}
    \newcommand*{\compute}              {\textsf{Compute}\xspace}
    \newcommand*{\decode}               {\textsf{Decode}\xspace}
    \newcommand*{\decrypt}              {\textsf{Dec}\xspace}
    \newcommand*{\encode}               {\textsf{Encode}\xspace}
    \newcommand*{\encrypt}              {\textsf{Enc}\xspace}
    \newcommand*{\eval}                 {\textsf{Eval}\xspace}
    \newcommand*{\keygen}               {\textsf{KeyGen}\xspace}
    \newcommand*{\probgen}              {\textsf{ProbGen}\xspace}
    
    \newcommand*{\verify}               {\textsf{Verify}\xspace}
    \newcommand*{\setup}                {\textsf{Setup}\xspace}

    \newcommand*{\threshold}[2]         {\ensuremath{#1}-out-of-\ensuremath{#2}-threshold\xspace}

        \newcommand*{\circlePercentage}[1]{
            \begin{tikzpicture}[scale=0.1]
                \draw (0,0) circle (1);
                \fill[fill=black] (0,0) -- (90:1) arc (90:90-#1*3.6:1) -- cycle;
            \end{tikzpicture}
        }
        \newcommand*{\closedcircle}{\circlePercentage{100}}
        \newcommand*{\halfopencircle}{\circlePercentage{50}}
        \newcommand*{\opencircle}{
            \begin{tikzpicture}[scale=0.1]
                \draw (0,0) circle (1);
            \end{tikzpicture}
        }

    \newcommand*{\NP}                   {\textsf{NP}\xspace}

    \newcommand*{\evaluationKey}        {\ensuremath{ek}\xspace}

    \newcommand*{\publicKey}            {\ensuremath{pk}\xspace}
    \newcommand*{\publicParameters}     {\textsf{pp}\xspace}

    \newcommand*{\secretKey}            {\ensuremath{sk}\xspace}
    \newcommand*{\securityParameter}    {\ensuremath{\kappa}\xspace}
    \newcommand*{\secret}               {\ensuremath{s}\xspace}
    \newcommand*{\verificationKey}      {\ensuremath{vk}\xspace}
    
    \newcommand*{\zkproof}              {\ensuremath{\pi}\xspace} 

    \theoremstyle{definition}
    \newtheorem{definition}{Definition}[section]

    \newcommand\rot[1]{\rlap{\rotatebox{45}{#1}}}

\begin{document}

    \begin{acronym}[zk-SNARK]
    \acro{ABE}                          {attribute-based encryption}
    \acro{CA}                           {Cochran-Armitage}
    \acro{CRH}                          {cryptographic hash function}
    \acro{DLT}                          {distributed ledger technology}
        \acrodefplural{DLT}                 {distributed ledger technologies}
    \acro{DP}                           {differential privacy}
    \acro{DPC}                          {decentralized private computation}
    \acro{DS}                           {disease susceptibility}
    \acro{HE}                           {homomorphic encryption}
        \acrodefindefinite{HE}              {an}{a}
    \acro{FHE}                          {fully homomorphic encryption}
        \acrodefindefinite{FHE}             {an}{a}
    \acro{FL}                           {federated learning}
        \acrodefindefinite{FL}              {an}{a}
    \acro{FS}                           {Fiat-Shamir}
    \acro{GC}                           {garbled circuit}
    \acro{GWAS}                         {Genome-Wide Association Study}
        \acrodefplural{GWAS}[GWASes]        {Genome-Wide Association Studies}
    \acro{HE}                           {homomorphic encryption}
    \acro{HVZK}                         {honest verifier zero-knowledge}
        \acrodefindefinite{HE}              {an}{a}
    \acro{LD}                           {linkage disequilibrium}
        \acrodefplural{LD}                  {linkage disequilibria}
    \acro{LDP}                          {local differential privacy}
    \acro{LHE}                          {leveled homomorphic encryption}
        \acrodefindefinite{LHE}             {an}{a}
    \acro{LSSS}                         {linear secret sharing scheme}
        \acrodefindefinite{LSSS}            {an}{a}
    \acro{LWE}                          {learning with errors}
        \acrodefindefinite{LWE}             {an}{a}
    \acro{MAC}                          {message authentication code}
        \acro{MAF}                          {minor allele frequency}
    \acrodefplural{MAF}                 {minor allele frequencies}
    \acro{ML}                           {machine learning}
    \acro{MPC}                          {secure multiparty computation}
        \acrodefindefinite{MPC}             {an}{a}
    \acro{NIZK}                         {non-interactive zero-knowledge proof}
    \acro{OT}                           {oblivious transfer}
    \acro{ORAM}                         {oblivious RAM}
    \acro{PET}                          {privacy-enhancing technology}
        \acrodefplural{PET}                 {privacy-enhancing technologies}
    \acro{PHE}                          {partially homomorphic encryption}
    \acro{PIR}                          {private information retrieval}
    \acro{PoR}                          {proof of retrievability}
        \acrodefplural{PoR}             {proofs of retrievability}
    \acro{PPC}                          {privacy-preserving computation}
    \acro{ppt}[p.p.t.\@]                {probabilistic polynomial time}
    \acro{PQ}                           {post-quantum}
    \acro{PSI}                          {private set intersection}
    \acro{SNP}                          {single nucleotide polymorphism}
    \acro{SNARK}                        {succinct non-interactive argument of knowledge}
        \acrodefplural{SNARK}               {succinct non-interactive arguments of knowledge}
    \acro{STARK}                        {succinct transparent argument of knowledge}
        \acrodefplural{STARK}               {succinct transparent arguments of knowledge}
    \acro{SSI}                          {self-sovereign identity}
        \acrodefindefinite{SSI}             {an}{a}
    \acro{SWHE}                         {somewhat homomorphic encryption}
        \acrodefindefinite{SWHE}            {an}{a}
    \acro{TEE}                          {trusted execution environment}
    \acro{TTP}                          {trusted third party}
    \acro{VC}                           {verifiable computing}
    \acro{VPPC}                         {verifiable, privacy-preserving computation}
    \acro{ZKP}                          {zero-knowledge proof}
    \acro{ZKPoK}                        {zero-knowledge proof of knowledge}
        \acrodefplural{ZKPoK}               {zero-knowledge proofs of knowledge}
    \acro{zk-SNARK}                     {zero-knowledge succinct non-interactive argument of knowledge}
        \acrodefplural{zk-SNARK}            {zero-knowledge succinct non-interactive arguments of knowledge}
    \acro{zk-STARK}                     {zero-knowledge succinct transparent argument of knowledge}
        \acrodefplural{zk-STARK}            {zero-knowledge succinct transparent arguments of knowledge}
\end{acronym}

    \title{Verifiable Privacy-Preserving Computing}

    \author{Tariq Bontekoe}
    \orcid{0000-0002-5331-4033}
    \affiliation{%
        \institution{University of Groningen}
        \city{Groningen}
        \country{the Netherlands}}
    \email{t.h.bontekoe@rug.nl}

    \author{Dimka Karastoyanova}
    \orcid{0000-0002-8827-2590}
    \affiliation{%
        \institution{University of Groningen}
        \city{Groningen}
        \country{the Netherlands}}
    \email{d.karastoyanova@rug.nl}

    \author{Fatih Turkmen}
    \orcid{0000-0002-6262-4869}
    \affiliation{%
        \institution{University of Groningen}
        \city{Groningen}
        \country{the Netherlands}}
    \email{f.turkmen@rug.nl}

    \begin{abstract}
        \Ac{PPC} methods, such as \ac{MPC} and \ac{HE}, are deployed increasingly often to guarantee data confidentiality in computations over private, distributed data.
        Similarly, we observe a steep increase in the adoption of \acp{ZKP} to guarantee (public) verifiability of locally executed computations.
        We project that applications that are data intensive and require strong privacy guarantees, are also likely to require verifiable correctness guarantees, especially when outsourced.
        While the combination of methods for verifiability and privacy protection has clear benefits, certain challenges stand before their widespread practical adoption.

        In this work, we analyze existing solutions that combine verifiability with privacy-preserving computations over distributed data, in order to preserve confidentiality and guarantee correctness at the same time.
        We classify and compare 37 different schemes, regarding solution approach, security, efficiency, and practicality.
        Lastly, we discuss some of the most promising solutions in this regard, and present various open challenges and directions for future research.
    \end{abstract}


    \maketitle

    \section{Introduction}\label{sec:introduction}
    In recent years, \ac{PPC} methods have played a significant role in enabling the computation of a function by two or more mutually distrusting parties over their respective private datasets.
    \ac{PPC} can be used to (effectively) enrich, diversify, or enlarge the available dataset and thereby obtain improved, more representative, or otherwise impossible results from a wide range of computations, e.g., statistics, \ac{ML}, or risk modeling.

    Due to their potential to allow for such collaborations without losing confidentiality, \ac{PPC} methods have been adopted in many domains, from healthcare~\cite{b.SurveyGenomicData2021}, to auctions~\cite{alvarezComprehensiveSurveyPrivacypreserving2020}, and finance~\cite{baumSoKPrivacyEnhancingTechnologies2023}.
    Due to recent advances in \ac{PPC} methods, we even observe usage of \ac{PPC} for data-intensive applications such as neural network inference~\cite{zamaZamaFullyHomomorphic2023}.

    Similarly, (homomorphic) \acp{MAC}, \acp{TEE}, and \acp{ZKP} allow honest parties to verify computations executed by other untrusted parties.
    These techniques can be applied when parties are mutually distrusting, or when auditability by external parties or the general public is required.
    Most notably, the introduction of efficient \acp{ZKP} has led to a wide range of \ac{VC} applications, ranging from verifiable outsourcing~\cite{jakobsenFrameworkOutsourcingSecure2014, schoenmakersTrinocchioPrivacyPreservingOutsourcing2016} to anonymous payment systems~\cite{ben-sassonZerocashDecentralizedAnonymous2014, bontekoeBalancingPrivacyAccountability2022}, and publicly verifiable voting~\cite{ramchenUniversallyVerifiableMPC2019, leeSAVERSNARKfriendlyAdditivelyhomomorphic2019}.

    Whilst \ac{PPC} methods almost always provide clear guarantees regarding privacy of data and/or computation within a certain security model, they often do not, by themselves, guarantee data authenticity or --- more importantly --- offer verifiability of computation.
    Recent advances, in both verifiability and \ac{PPC}, have led to schemes that combine a \ac{PPC} method with a certain verifiability approach to enable \acp{VPPC} between distrusting parties.
    These solutions are particularly relevant when not only input privacy, but also the correctness of a computation's results are at stake, e.g., computation outsourcing~\cite{schoenmakersTrinocchioPrivacyPreservingOutsourcing2016}, decentralized applications~\cite{boweZEXEEnablingDecentralized2020}, or e-voting~\cite{panjaSecureEndtoendVerifiable2021}.

    The combination of privacy-preserving and verifiable computations is needed in a wide range of settings, especially when private data is distributed over mutually distrusting parties.
    In the situation where a group of parties wants to collaboratively compute a function over their private data, but do not have sufficient computational resources to do so, outsourcing of the computation (to the cloud) has become a common practice.
    However, in addition to privacy concerns, the parties have no guarantee whether the computational results are actually correct.
    By using \acp{VPPC}, all parties can be sure that their data remains private ánd their results are correct, thereby increasing trust in outsourced computations.

    An even more prolific setting these days, is where an external party or the general \emph{public} needs to be able to verify the results of a privacy-preserving computation.
    In electronic voting, auctions, and many blockchain/\ac{DLT} applications, the public, i.e., a group of people that need not be known a priori, should be able to verify correctness to ascertain honest election results, auction outcomes, and so on.

    Due to the diversity in both \ac{PPC} and verifiability techniques, we observe a wide range of \ac{VPPC} schemes, making analyses and comparisons a non-trivial task.
    In this work, we aim to answer the following questions to better understand the differences between different \ac{VPPC} schemes and how they can be applied in practice:

    \begin{enumerate}
        \item Which classes of \ac{VPPC} schemes are there?
        In which use cases or settings are they most applicable?
        \item Under which conditions do privacy and verifiability hold?
        Which schemes provide public verifiability?
        \item How efficient/suitable are the schemes for practical usage?
        Are there any other limiting factors?
        \item Which open problems are still to be solved?
        How can current solutions be improved upon?
    \end{enumerate}

    \subsection*{Contributions}
    In this work, we first determine the most interesting \ac{VPPC} schemes and classify them into four main classes.
    This is followed by a high-level comparison and discussion of underexposed topics.
    Next, we analyze and compare the different methods, by dividing the solutions per construction method.
    Finally, we describe open challenges for the different solutions and discuss which constructions are most promising.
    To summarize, our contributions are as follows:
    \begin{itemize}
        \item We identify and classify 37 existing solutions from academic literature into four main classes.
        We further divide each class based on the schemes' distinguishing primitives.
        \item We compare these works based on the settings in which privacy and verifiability are provided.
        Moreover, we study the efficiency of the different schemes, and discuss practical aspects such as public verifiability and use case.
        \item We compare the different construction methods for different \ac{VPPC} schemes and show which ones seem most promising for practical usage.
        Next to this, we denote open challenges and improvement areas for the different solutions, regarding security, efficiency, or practicality.
    \end{itemize}

    \subsection*{Organization}
    The remainder of this work is organized as follows.
    \cref{sec:preliminaries} discusses preliminaries regarding verifiability approaches and \cref{sec:privacy-preserving-computations-on-distributed-data} discusses relevant background information regarding (non-verifiable) \acp{PPC}.
    We classify the \ac{VPPC} schemes we found in \cref{sec:classification} and present formal definitions and a high-level comparison in \cref{sec:verifiable-privacy-preserving-computation}.
    Each solution (paradigm) and its challenges are discussed in detail, per class, in \cref{sec:dlt-based-vppc,sec:mpc-based-vppc,sec:he-based-vppc,sec:dp-based-vppc}.
    We conclude in \cref{sec:conclusion}.

    \section{Preliminaries}\label{sec:preliminaries}
    This section provides relevant background on the three methods for (public) verifiability that are used to construct the significant majority of \ac{VPPC} schemes: \acp{ZKP}, homomorphic \acp{MAC} and \acp{TEE}.

    \subsection{Zero-knowledge proofs}\label{subsec:zero-knowledge-proofs}
    A \acl{ZKP} allows a \emph{prover} to prove the truth value of a certain statement to a (group of) \emph{verifier}(s), whilst hiding all private information~\cite{goldwasserKnowledgeComplexityInteractive1989}.
    This is useful in cases where the truth value of a claim is not evident, but the prover does hold private information that is sufficient to create a proof thereof.
    \Iac{ZKP} scheme should at least satisfy: \emph{completeness}, \emph{soundness}, and \emph{zero-knowledgeness}~\cite{blumNoninteractiveZeroknowledgeIts1988}.

    Initially, most \ac{ZKP} schemes were solutions for specific problems, mostly in the form of $\Sigma$-protocols, e.g., Schnorr's protocol~\cite{schnorrEfficientIdentificationSignatures1990}.
    While these schemes could be used to create proofs for arbitrary \NP-statements, they often have a proof size and verification time that scale linearly (or worse) in the computation size.
    On top of that, $\Sigma$-protocols are \emph{interactive} which is often undesired.
    On the plus side, $\Sigma$-protocols can be made \emph{non-interactive} using the \ac{FS} heuristic~\cite{fiatHowProveYourself1987}, which additionally enables \emph{public verifiability}.

    The introduction of Pinocchio~\cite{parnoPinocchioNearlyPractical2013} as the first \emph{succinct \ac{ZKP}} or \ac{zk-SNARK} gave the first scheme with efficient verification and small proof size at the cost of requiring a \emph{trusted setup}.
    Moreover, it could be used to prove any computation that can be described as an arithmetic circuit, i.e., it supports \ac{VC}.
    Pinocchio was followed by many schemes with improved efficiency or different properties, e.g.,
    \textit{Groth16}~\cite{grothSizePairingBasedNoninteractive2016}, \textit{Marlin}~\cite{chiesaMarlinPreprocessingZkSNARKs2020}, \textit{Fractal}~\cite{chiesaFractalPostquantumTransparent2020}.

    Another line of efficient \acp{NIZK} was started with the introduction of Bulletproofs~\cite{bunzBulletproofsShortProofs2018}.
    More recently, the efficient construction used for Bulletproofs has been applied to create $\Sigma$-bullets~\cite{bunzZetherPrivacySmart2020}, or compressed $\Sigma$-protocol theory~\cite{attemaCompressedSProtocolTheory2020}.
    The protocols based on the latter can be used to construct \acp{ZKP} for arithmetic circuits from standard security assumptions, which is not the case for \acp{zk-SNARK}, at the cost of an increased proof size and verification time.

    For a more extensive and up-to-date overview of \ac{ZKP} schemes we refer the reader to, e.g.,~\cite{thalerProofsArgumentsZeroKnowledge2022,zkproofZKProofWikiConcrete2022}.

    \subsection{Homomorphic MACs}\label{subsec:homomorphic-macs}
    A \acf{MAC} scheme enables a user to generate a short, unforgeable tag for a message such that, later, any recipient of both tag and message can verify the integrity and authenticity of that message.
    The tag is computed using a secret key and is verified by checking the \ac{MAC} against this same secret authentication key.
    Regular \acp{MAC} are non-malleable on purpose, i.e., it is undesirable for parties to be able to alter both message and tag in such a way that verification still succeeds.
    However, homomorphic \acp{MAC} deliberately allow for a restricted and pre-defined set of operations to be applied to both message and tag such that verification is still successful.

    \subsection{Trusted Execution Environments}\label{subsec:trusted-execution-environments}
    \Acp{TEE} are dedicated hardware (and software) components, running in an isolated part of the main CPU, that are capable of running code while maintaining input privacy and integrity.
    Nowadays, \acp{TEE} are offered by most major hardware vendors as well as a number of open source projects~\cite{intelIntelSoftwareGuard2023,amdAMDSecureEncrypted2023,enarxEnarxConfidentialComputing2023}.
    Code running on \iac{TEE} is isolated in such a way that it cannot be tampered with by any other process.

    A user wishing to use a remote \ac{TEE} securely can verify that it is running the right code and has been created using safe setup procedures, by using a process known as (remote) attestation, as supported by, e.g., Intel SGX~\cite{intelIntelSoftwareGuard2023}.
    However, the user does have to trust that the hardware has not been tampered with or is broken in an undetectable manner.
    There are cases in academic literature where \ac{TEE} security has been broken, however due to the novelty of the technology it is still unclear to which extent such attacks are possible on the most recent \ac{TEE} hardware~\cite{munozSurveySecurityTrusted2023}.

    \section{Privacy-preserving computations on distributed data}\label{sec:privacy-preserving-computations-on-distributed-data}
    There are different technologies that can be used to construct schemes for \ac{PPC} over distributed data.
    Some are specifically suitable as building blocks for \acp{VPPC}: \acs{HE}, \acs{MPC}, \acs{DLT}, and \ac{DP}.
    We will see in \cref{subsec:high-level-comparison} that different building blocks are best suitable for different use cases.

    \subsection{Homomorphic encryption}\label{subsec:homomorphic-encryption}
    \Ac{HE} denotes a collection of (asymmetric) encryption schemes that allow users to perform operations on encrypted data without decrypting it first.
    In general, \iac{HE} scheme is described by four \ac{ppt} algorithms $(\keygen, \encrypt, \decrypt, \eval)$, respectively the key generation, encryption, decryption, and function evaluation algorithm.

    While there are different ways to represent the function that is to be evaluated on \ac{HE} ciphertexts, most constructions translate it to an arithmetic circuit.
    We distinguish, following~\cite{armknechtGuideFullyHomomorphic2015}, several types of homomorphic schemes in increasing order of functionality.

    \emph{\Ac{PHE}} is a type of homomorphic encryption that is only \emph{additively} or only \emph{multiplicatively} homomorphic, implying that it can only evaluate arithmetic circuits consisting solely of addition, respectively multiplication gates.

    \emph{\Ac{SWHE}} and \emph{\ac{LHE}} constructions can evaluate arithmetic circuits consisting of both addition and multiplication gates.
    However, the circuits that can be evaluated may be limited in the number of operations (of either type).
    These schemes can also have practical limitations, such as ciphertexts or key sizes that increase exponentially with the number of multiplications.

    \emph{\Ac{FHE}} schemes can evaluate arbitrary arithmetic circuits, without exponential blowup of ciphertext or key size.
    This generality often comes at the cost of computationally expensive bootstrapping after a number of multiplications~\cite{gentryFullyHomomorphicEncryption2009}.
    We observe that bootstrapping is becoming more practical in recent years, as exemplified by, e.g., TFHE~\cite{chillottiTFHEFastFully2020}.

    In the remainder of this work, when we talk about \ac{HE} we only refer to \ac{SWHE}, \ac{LHE} and predominantly \ac{FHE} schemes.

    \subsection{Secure multiparty computation}\label{subsec:secure-multiparty-computation}
    \Ac{MPC} is a collection of techniques that allows for performing joint computations over distributed data, while keeping all input data private and revealing nothing but the computation output.
    There are several ways to construct \iac{MPC} protocol or framework.
    We will discuss the most common constructions.

    Generally, there are $n$ parties that participate in \iac{MPC} computation, and any participating party contributes with their respective private inputs.
    The $n$ parties jointly execute an \emph{interactive} protocol to compute the function output from their private inputs.
    Some of these parties might be corrupted by an adversary.
    Therefore, any secure \ac{MPC} protocol should at least satisfy: \emph{input privacy}, \emph{correctness}, and \emph{independence of inputs}~\cite{lindellSecureMultipartyComputation2020}.
    Moreover, in the active security setting (see \cref{sec:threat-models}), the scheme should also be \textit{secure} against a certain number of malicious parties, meaning that correctness should hold when these parties deviate from the protocol.

    In \cref{sec:background-mpc} we summarize the most frequently used constructions for \ac{MPC}: \emph{secret sharing}, \emph{\acp{GC}}, and \emph{\ac{PHE}-based}.

    \subsection{Distributed ledger technologies}\label{subsec:distributed-ledger-technologies}
    \Acp{DLT} cover a collection of technologies where a (large) number of parties collectively create, maintain, and agree on a shared state or ledger.
    The shared state can be known completely to all parties, but can also be distributed over the respective parties.
    A well-known example of \iac{DLT} is blockchain, where all parties agree --- through a consensus mechanism --- on a public ledger describing the full system state.
    Blockchain is best known for realizing decentralized currencies such as Bitcoin~\cite{nakamotoBitcoinPeertoPeerElectronic2008}, or smart contract systems like Ethereum~\cite{woodEthereumSecureDecentralised2014}.
    However, \acp{DLT}, also have applications in many other domains: from identity management~\cite{ahmedBlockchainBasedIdentityManagement2022} (e.g., \acl*{SSI}) to healthcare~\cite{adanurdedeturkBlockchainGenomicsHealthcare2021}.
    We specifically observe the proliferation of decentralized apps, or \emph{dApps}, each of which can be seen as a blockchain-based autonomous application, whose behavior is defined by a collection of scripts, or \emph{smart contracts}.

    In lockstep with an increasing number of dApps, we observe an increase in using \ac{DLT} for decentralized computations.
    Such decentralized computations can be used in any situation where multiple parties wish to perform a group computation.
    A group computation can be: (1) a single computation on distributed data; (2) a decentralized computation on centrally stored data; or (3) a combination of the two.
    Especially of interest for this work are decentralized computations with input privacy, e.g.,~\cite{bernalbernabePrivacyPreservingSolutionsBlockchain2019,alghazwiBlockchainGenomicsSystematic2022}.

    \subsection{Differential Privacy}\label{subsec:differential-privacy}
    In \ac{DP} methods, sensitive input data, or results obtained from sensitive data, are perturbed such that the underlying input data is hidden.
    The technique was first formalized in~\cite{dworkDifferentialPrivacy2006} and while it gives a weaker privacy guarantee than is provided by, e.g., \ac{MPC} or \ac{HE} it also comes with benefits.
    \ac{DP} computations are often significantly more efficient than other methods, and the fact that results are not exact, due the added perturbation, can be used to prevent other types of attacks such as deanonymization using external sources, or inference attacks on \ac{ML} models~\cite{abadiDeepLearningDifferential2016,chenDifferentialPrivacyProtection2021}.

    \ac{DP} is often divided in two settings: \emph{\ac{LDP}} and a model with a \emph{trusted curator}.
    In \ac{LDP}, the data owners perturb the data themselves (locally), and subsequently send their \ac{DP}-data to an (untrusted) curator who computes the final result.
    For the model with a trusted curator, the parties send their differentially private data directly to the trusted curator, who computes the final result from the data.
    The final result is only released to the public after adding differentially private noise in order to protect the sensitive input data.

    There have been many variations on these models, including the more recent shuffle model~\cite{bittauProchloStrongPrivacy2017,cheuManipulationAttacksLocal2021}, combining \ac{DP} with \ac{MPC}~\cite{pettaiCombiningDifferentialPrivacy2015}, or verifiable \ac{DP}.
    The latter has a basis in cryptographic randomized response techniques as used in~\cite{ambainisCryptographicRandomizedResponse2004}, and can be used to realize, e.g., verifiable differentially private polling~\cite{munillagarridoVerifiableDifferentiallyPrivatePolling2022}.
    In the remainder of this work we primarily focus on verifiable \ac{DP} for generic applications.

    \section{Classification}\label{sec:classification}
    In this section, we first summarize the approach for our literature search and then classify the resulting works in four classes.
    This is followed by a description of the relevant properties that we have analyzed for each of the works included in our study.
    These properties are formally defined in \cref{sec:verifiable-privacy-preserving-computation}, followed by a comparison of the classes and a discussion of generic, high-level observations.

    \subsection{Literature search}\label{subsec:literature-search}
    To obtain a comprehensive list of relevant works, we first determined a small set of recent, relevant articles on \ac{VPPC}.
    Specifically, we determined at least one very recent \enquote*{seed} article for different underlying (privacy-preserving) computation techniques, by querying the Google Scholar and Scopus databases.
    The most recent, at the time of writing, matching verifiable \ac{MPC} paper~\cite{riviniusPubliclyAccountableRobust2022} was found with the query: \emph{(public verifiability OR public auditability) AND (MPC OR SMPC OR multi-party computation OR secure multi-party computation)}.
    Two relevant, recent verifiable \ac{HE} papers~\cite{viandVerifiableFullyHomomorphic2023, louVFHEVerifiableFully2023} were found using: \emph{(verifiable OR public verifiability OR public auditability) AND (homomorphic encryption OR fully homomorphic encryption)}.
    The most recent, at the time of writing, matching verifiable \ac{DP} paper~\cite{movsowitzdavidowPrivacyPreservingTransactionsVerifiable2023} was found with the query: \emph{(verifiable OR public verifiability OR public auditability) AND (differential privacy)}.
    The final seed paper~\cite{boweZEXEEnablingDecentralized2020} on \ac{DLT}-based privacy-preserving computations was, up front, known to be recent and relevant, hence no specific query was used.

    Subsequently, we checked all papers in the reference list of these \enquote*{seed} articles to the second degree.
    Next to this, all articles that referenced our \enquote*{seed} articles were also checked.
    After filtering the articles on their relevance, based on title, abstract, and a quick scan of the content, we found 37 relevant works that present a novel or improved solution for \acp{VPPC}.

\defcitealias{baumPubliclyAuditableSecure2014}                              {\emph{BDO14}}
\defcitealias{schoenmakersUniversallyVerifiableMultiparty2015}              {\emph{SV15}}
\defcitealias{cunninghamCatchingMPCCheaters2016}                            {\emph{CFY16}}
\defcitealias{priviledgeprojectRevisionExtendedCore2021}                    {\emph{PRI21}}
\defcitealias{riviniusPubliclyAccountableRobust2022}                        {\emph{RRRK22}}
\defcitealias{duttaComputeVerifyEfficient2022}                              {\emph{DGPS22}}
\defcitealias{schoenmakersTrinocchioPrivacyPreservingOutsourcing2016}       {\emph{SVdV16}}
\defcitealias{veeningenPinocchioBasedAdaptiveZkSNARKs2017}                  {\emph{Vee17}}
\defcitealias{kanjalkarPubliclyAuditableMPCasaService2021}                  {\emph{KZGM21}}
\defcitealias{ozdemirExperimentingCollaborativeZkSNARKs2022}                {\emph{OB22}}
\defcitealias{jakobsenFrameworkOutsourcingSecure2014}                       {\emph{JNO14}}
\defcitealias{schabhuserFunctionDependentCommitmentsVerifiable2018}         {\emph{SBDB18}}
\defcitealias{ramchenUniversallyVerifiableMPC2019}                          {\emph{RCPT19}}
\defcitealias{baldimtsiCrowdVerifiableZeroKnowledge2020}                    {\emph{BKZZ20}}
\defcitealias{baumEfficientConstantRoundMPC2020}                            {\emph{BOSS20}}
\defcitealias{gennaroFullyHomomorphicMessage2013}                           {\emph{GW13}}
\defcitealias{catalanoPracticalHomomorphicMACs2013}                         {\emph{CF13}}
\defcitealias{fioreEfficientlyVerifiableComputation2014}                    {\emph{FGP14}}
\defcitealias{liPrivacyPreservingHomomorphicMACs2018}                       {\emph{LWZ18}}
\defcitealias{chatelVerifiableEncodingsSecure2022}                          {\emph{CKPH22}}
\defcitealias{fioreBoostingVerifiableComputation2020}                       {\emph{FNP20}}
\defcitealias{boisFlexibleEfficientVerifiable2020}                          {\emph{BCFK20}}
\defcitealias{ganeshRinocchioSNARKsRing2021}                                {\emph{GNS21}}
\defcitealias{viandVerifiableFullyHomomorphic2023}                          {\emph{VKH23}}
\defcitealias{natarajanCHEXMIXCombiningHomomorphic2021}                     {\emph{NLDD21}}
\defcitealias{louVFHEVerifiableFully2023}                                   {\emph{Lou+23}}
\defcitealias{gennaroNoninteractiveVerifiableComputing2010}                 {\emph{GGP10}}
\defcitealias{kosbaHawkBlockchainModel2016}                                 {\emph{Kos+16}}
\defcitealias{boweZEXEEnablingDecentralized2020}                            {\emph{Bow+20}}
\defcitealias{bunzZetherPrivacySmart2020}                                   {\emph{BAZB20}}
\defcitealias{chengEkidenPlatformConfidentialityPreserving2019}             {\emph{Che+19}}
\defcitealias{narayanVerifiableDifferentialPrivacy2015}                     {\emph{NFPH15}}
\defcitealias{movsowitzdavidowPrivacyPreservingTransactionsVerifiable2023}  {\emph{MMT23}}
\defcitealias{tsaloliDifferentialPrivacyMeets2023}                          {\emph{TM23}}
\defcitealias{biswasVerifiableDifferentialPrivacy2023}                      {\emph{BC23}}
\defcitealias{katoPreventingManipulationAttack2021}                         {\emph{KCY21}}
\defcitealias{cuvelierVerifiableMultipartyComputation2016}                  {\emph{CP16}}

\begin{table*}
    \footnotesize
    \centering
    \caption{Classification of \ac{VPPC} schemes$^\dagger$.}
    \label{tab:classification}
    \begin{threeparttable}
        \begin{tabular}{lllllllllllll} \toprule
            Name                                                                     & \rot{Paper}                                                        & \rot{Class} & \rot{Verifiability paradigm}       & \rot{Input privacy}        & \rot{Security} & \rot{Public verifiability} & \rot{Experimental evaluation} & \rot{Implementation available} & \rot{Assumptions}            & \rot{Other}                                              \\ \midrule
            \citetalias{baumPubliclyAuditableSecure2014}                             & \cite{baumPubliclyAuditableSecure2014}                             & \acs*{MPC}  & non-succinct \ac{ZKP}              & 1H                         & \closedcircle  & \closedcircle              & \opencircle                   & \opencircle                    & Standard                     & -                                                        \\
            \citetalias{schoenmakersUniversallyVerifiableMultiparty2015}             & \cite{schoenmakersUniversallyVerifiableMultiparty2015}             &             & non-succinct \ac{ZKP}              & 1H                         & \closedcircle  & \closedcircle              & \opencircle                   & \opencircle                    & Standard                     & -                                                        \\
            \citetalias{cunninghamCatchingMPCCheaters2016}                           & \cite{cunninghamCatchingMPCCheaters2016}                           &             & non-succinct \ac{ZKP}              & 1H                         & \closedcircle  & \closedcircle              & \opencircle                   & \opencircle                    & Standard                     & $(+)$ Optimistic                                         \\
            \citetalias{priviledgeprojectRevisionExtendedCore2021}                   & \cite{priviledgeprojectRevisionExtendedCore2021}                   &             & non-succinct \ac{ZKP}              & HM                         & \closedcircle  & \closedcircle              & \opencircle                   & \closedcircle                  & Standard                     & -                                                        \\
            \citetalias{riviniusPubliclyAccountableRobust2022}                       & \cite{riviniusPubliclyAccountableRobust2022}                       &             & non-succinct \ac{ZKP}              & T                          & \closedcircle  & \closedcircle              & \closedcircle                 & \opencircle                    & Standard                     & $(+)$ Robust; Without robustness, input privacy is 1H    \\
            \citetalias{duttaComputeVerifyEfficient2022}                             & \cite{duttaComputeVerifyEfficient2022}                             &             & non-succinct \ac{ZKP}              & T                          & T              & \opencircle                & \closedcircle                 & \opencircle                    & Standard                     & $(+)$ Authenticated inputs                               \\
            \citetalias{schoenmakersTrinocchioPrivacyPreservingOutsourcing2016}      & \cite{schoenmakersTrinocchioPrivacyPreservingOutsourcing2016}      &             & succinct \ac{ZKP}                  & HM                         & \closedcircle  & \closedcircle              & \closedcircle                 & \opencircle                    & One-time, trusted setup      & -                                                        \\
            \citetalias{veeningenPinocchioBasedAdaptiveZkSNARKs2017}                 & \cite{veeningenPinocchioBasedAdaptiveZkSNARKs2017}                 &             & succinct \ac{ZKP}                  & T                          & \closedcircle  & \closedcircle              & \closedcircle                 & \closedcircle                  & Adaptive, trusted setup      & -                                                        \\
            \citetalias{kanjalkarPubliclyAuditableMPCasaService2021}                 & \cite{kanjalkarPubliclyAuditableMPCasaService2021}                 &             & succinct \ac{ZKP}                  & T                          & \closedcircle  & \closedcircle              & \closedcircle                 & \closedcircle                  & Universal, trusted setup     & -                                                        \\
            \citetalias{ozdemirExperimentingCollaborativeZkSNARKs2022}               & \cite{ozdemirExperimentingCollaborativeZkSNARKs2022}               &             & succinct \ac{ZKP}                  & T                          & \closedcircle  & \closedcircle              & \closedcircle                 & \closedcircle                  & Same as \acs*{zk-SNARK} used & -                                                        \\
            \citetalias{jakobsenFrameworkOutsourcingSecure2014}                      & \cite{jakobsenFrameworkOutsourcingSecure2014}                      &             & \acp{MAC}                          & 1H                         & 1H             & \opencircle                & \opencircle                   & \opencircle                    & Standard                     & -                                                        \\
            \citetalias{schabhuserFunctionDependentCommitmentsVerifiable2018}        & \cite{schabhuserFunctionDependentCommitmentsVerifiable2018}        &             & function-dep.\ comm.\              & T                          & \closedcircle  & \closedcircle              & \opencircle                   & \opencircle                    & Standard                     & $(-)$ Only linear computations                           \\
            \citetalias{ramchenUniversallyVerifiableMPC2019}                         & \cite{ramchenUniversallyVerifiableMPC2019}                         &             & verifiable key switching           & 1H                         & \closedcircle  & \closedcircle              & \closedcircle                 & \closedcircle                  & Standard                     & -                                                        \\
            \citetalias{baldimtsiCrowdVerifiableZeroKnowledge2020}                   & \cite{baldimtsiCrowdVerifiableZeroKnowledge2020}                   &             & crowd-verifiable \ac{ZKP}          & 1H                         & \closedcircle  & \opencircle                & \opencircle                   & \opencircle                    & Standard                     & $(-)$ Interactive                                        \\
            \citetalias{baumEfficientConstantRoundMPC2020}                           & \cite{baumEfficientConstantRoundMPC2020}                           &             & commitment + hash                  & 1H                         & 1H             & \closedcircle              & \opencircle                   & \opencircle                    & Standard                     & -                                                        \\ \midrule
            \citetalias{gennaroFullyHomomorphicMessage2013}                          & \cite{gennaroFullyHomomorphicMessage2013}                          & \acs*{HE}   & homomorphic \ac{MAC}               & \closedcircle              & \closedcircle  & \opencircle                & \opencircle                   & \opencircle                    & Standard                     & $(-)$ Inefficient verification                           \\
            \citetalias{catalanoPracticalHomomorphicMACs2013}                        & \cite{catalanoPracticalHomomorphicMACs2013}                        &             & homomorphic \ac{MAC}               & \closedcircle              & \closedcircle  & \opencircle                & \opencircle                   & \opencircle                    & Standard                     & $(-)$ Only bounded depth arithmetic circuits             \\
            \citetalias{fioreEfficientlyVerifiableComputation2014}                   & \cite{fioreEfficientlyVerifiableComputation2014}                   &             & homomorphic \ac{MAC}               & \closedcircle              & \closedcircle  & \opencircle                & \closedcircle                 & \opencircle                    & Standard                     & $(-)$ Quadratic circuits only                            \\
            \citetalias{liPrivacyPreservingHomomorphicMACs2018}                      & \cite{liPrivacyPreservingHomomorphicMACs2018}                      &             & homomorphic \ac{MAC}               & \closedcircle              & \closedcircle  & \opencircle                & \opencircle                   & \opencircle                    & Standard                     & $(-)$ Unclear how to do other ops.\ than add and multiply\\
            \citetalias{chatelVerifiableEncodingsSecure2022}                         & \cite{chatelVerifiableEncodingsSecure2022}                         &             & homomorphic \ac{MAC}               & \closedcircle              & \closedcircle  & \opencircle                & \closedcircle                 & \opencircle                    & Standard                     & -                                                        \\
            \citetalias{fioreBoostingVerifiableComputation2020}                      & \cite{fioreBoostingVerifiableComputation2020}                      &             & succinct \ac{ZKP}                  & \closedcircle              & \closedcircle  & \closedcircle              & \opencircle                   & \opencircle                    & One-time, trusted setup      & -                                                        \\
            \citetalias{boisFlexibleEfficientVerifiable2020}                         & \cite{boisFlexibleEfficientVerifiable2020}                         &             & succinct \ac{ZKP}                  & \closedcircle              & \closedcircle  & \closedcircle              & \opencircle                   & \opencircle                    & Same as \acs*{zk-SNARK} used & $(-)$ Only logspace-uniform circuits                     \\
            \citetalias{ganeshRinocchioSNARKsRing2021}                               & \cite{ganeshRinocchioSNARKsRing2021}                               &             & succinct \ac{ZKP}                  & \closedcircle              & \closedcircle  & \closedcircle              & \opencircle                   & \opencircle                    & One-time, trusted setup      & $(+)$ Also has designated-verifier version               \\
            \citetalias{viandVerifiableFullyHomomorphic2023}                         & \cite{viandVerifiableFullyHomomorphic2023}                         &             & succinct \ac{ZKP}                  & \closedcircle              & \closedcircle  & \closedcircle              & \closedcircle                 & \closedcircle                  & Same as \acs*{zk-SNARK} used & -                                                        \\
            \citetalias{natarajanCHEXMIXCombiningHomomorphic2021}                    & \cite{natarajanCHEXMIXCombiningHomomorphic2021}                    &             & \acs*{TEE}                         & \closedcircle              & \closedcircle  & \opencircle                & \closedcircle                 & \closedcircle                  & Trusted hardware             & -                                                        \\
            \citetalias{louVFHEVerifiableFully2023}                                  & \cite{louVFHEVerifiableFully2023}                                  &             & blind hash                         & \closedcircle              & \closedcircle  & \opencircle                & \closedcircle                 & \opencircle                    & No security analysis         & $(-)$ Unclear how to do other ops.\ than matrix mult.\   \\
            \citetalias{gennaroNoninteractiveVerifiableComputing2010}                & \cite{gennaroNoninteractiveVerifiableComputing2010}                &             & adapted Yao's \acs*{GC}            & \closedcircle              & \closedcircle  & \opencircle                & \opencircle                   & \opencircle                    & Standard                     & $(-)$ One-time setup for each function                   \\ \midrule
            \citetalias{kosbaHawkBlockchainModel2016}                                & \cite{kosbaHawkBlockchainModel2016}                                & \acs*{DLT}  & succinct \ac{ZKP}                  & \halfopencircle            & \closedcircle  & \closedcircle              & \closedcircle                 & \opencircle                    & Same as \acs*{zk-SNARK} used & $(-)$ Trusted computation party w.r.t.\ input privacy    \\
            \citetalias{boweZEXEEnablingDecentralized2020}                           & \cite{boweZEXEEnablingDecentralized2020}                           &             & succinct \ac{ZKP}                  & \halfopencircle            & \closedcircle  & \closedcircle              & \closedcircle                 & \closedcircle                  & Same as \acs*{zk-SNARK} used & $(+)$ Function privacy                                   \\
            \citetalias{bunzZetherPrivacySmart2020}                                  & \cite{bunzZetherPrivacySmart2020}                                  &             & non-succinct \ac{ZKP}              & \halfopencircle            & \closedcircle  & \closedcircle              & \closedcircle                 & \opencircle                    & Standard                     & $(-)$ Only has anonymous payments                        \\
            \citetalias{chengEkidenPlatformConfidentialityPreserving2019}            & \cite{chengEkidenPlatformConfidentialityPreserving2019}            &             & \acs*{TEE}                         & \closedcircle              & \closedcircle  & \opencircle                & \closedcircle                 & \closedcircle                  & Trusted hardware             & -                                                        \\ \midrule
            \citetalias{narayanVerifiableDifferentialPrivacy2015}                    & \cite{narayanVerifiableDifferentialPrivacy2015}                    & \acs*{DP}   & succinct \ac{ZKP}                  & \halfopencircle$^\ddagger$ & \closedcircle  & \closedcircle              & \closedcircle                 & \opencircle                    & Same as \acs*{zk-SNARK} used & $(-)$ Trusted curator; Analysts get access to input data \\
            \citetalias{movsowitzdavidowPrivacyPreservingTransactionsVerifiable2023} & \cite{movsowitzdavidowPrivacyPreservingTransactionsVerifiable2023} &             & succinct \ac{ZKP}                  & \closedcircle$^\ddagger$   & \closedcircle  & \closedcircle              & \closedcircle                 & \closedcircle                  & Same as \acs*{zk-SNARK} used & $(-)$ Computation done by the analyst is not verifiable  \\
            \citetalias{tsaloliDifferentialPrivacyMeets2023}                         & \cite{tsaloliDifferentialPrivacyMeets2023}                         &             & \ac{ZKP}                           & \closedcircle$^\ddagger$   & \closedcircle  & \closedcircle              & \opencircle                   & \opencircle                    & Same as \acs*{ZKP} used      & $(-)$ Trusted curator; Only described on high-level      \\
            \citetalias{biswasVerifiableDifferentialPrivacy2023}                     & \cite{biswasVerifiableDifferentialPrivacy2023}                     &             & non-succinct \ac{ZKP} + \acs*{MPC} & \closedcircle$^\ddagger$   & \closedcircle  & \closedcircle              & \closedcircle                 & \closedcircle                  & Standard                     & $(-)$ Trusted verifier needed for public verifiability   \\
            \citetalias{katoPreventingManipulationAttack2021}                        & \cite{katoPreventingManipulationAttack2021}                        &             & non-succinct \ac{ZKP}              & \closedcircle$^\ddagger$   & \closedcircle  & \opencircle                & \closedcircle                 & \closedcircle                  & Standard                     & $(-)$ Computation done by the analyst is not verifiable  \\ \midrule
            \citetalias{cuvelierVerifiableMultipartyComputation2016}                 & \cite{cuvelierVerifiableMultipartyComputation2016}                 & \acs*{TTP}  & non-succinct \ac{ZKP}              & \closedcircle              & \closedcircle  & \closedcircle              & \closedcircle                 & \closedcircle                  & Trusted third party          & -                                                        \\ \bottomrule
        \end{tabular}
        \begin{tablenotes}[para]
            \item[$\dagger$] 1H = yes, if 1 honest party; T = yes, if threshold; HM = yes, if honest majority; \closedcircle = yes; \opencircle = no; \halfopencircle = partial; $(+)$/$(-)$ = positive/negative aspect.
            \item[$\ddagger$] Differential privacy.
        \end{tablenotes}
    \end{threeparttable}
\end{table*}

    \subsection{VPPC classes}\label{subsec:vppc-classes}
    Based on our literature review, we divide the works we found into four main classes of \ac{VPPC} schemes, based on the underlying (privacy-preserving) computation technique.

    An overview of our classification is given in \cref{tab:classification}, including the names we use to refer to the specific schemes in the remainder of this work.
    This table also includes a subdivision of each class based on the distinguishing technique used to achieve verifiability, along with other properties.
    We consider the following classes:
    \begin{itemize}
        \item \emph{MPC-based VPPC}\quad This class gathers all solutions that rely on --- pre-existing, adapted, or novel --- \ac{MPC} schemes for computing functions on distributed private input data.
        \item \emph{HE-based VPPC}\quad This class gathers all solutions that rely on existing \ac{HE} for computing functions on private data.
        The private input data may be distributed over multiple parties, but this need not be the case, e.g., when considering verifiable outsourcing.
        \item \emph{DLT-based VPPC}\quad This class gathers all schemes that rely on \ac{DLT} for realizing computation and communication between the parties.
        For most solutions in this class, data is only kept private from outside observers that do not partake in the computation.
        \item \emph{DP-based VPPC}\quad This class gathers all solutions that rely on \ac{DP} for evaluating functions on private input data.
        \item \emph{TTP-based VPPC}\quad Finally, we consider \citetalias{cuvelierVerifiableMultipartyComputation2016} separately, since it solely relies on \iac{TTP} to guarantee input privacy.
        Whilst it does use \acp{ZKP} to achieve verifiability, we feel that the use of a single \ac{TTP} for input privacy, makes this scheme unsuitable for a real-world implementation of \iac{VPPC} scheme.
        For that reason, we do not further discuss this class.
    \end{itemize}

    Each of the remaining classes is discussed in detail in \cref{sec:dlt-based-vppc,sec:mpc-based-vppc,sec:he-based-vppc,sec:dp-based-vppc}, where we also describe the open challenges.
    A summary thereof is provided in \cref{tab:summary_paradigms,tab:summary_challenges}.
    We provide a brief overview of interesting techniques for works that are adjacent to \ac{VPPC} schemes, but cannot be classified as such, in \cref{sec:adjacent-solutions}.

    \subsection{Properties}\label{subsec:properties}
    Below, we describe the properties we consider in our analysis of the considered \ac{VPPC} schemes.

    \subsubsection{Security, privacy, and public verifiability}
    The first category of the properties we consider are related to security and verifiability, and are formally described in \cref{subsec:definitions}.
    We will not explicitly state whether schemes are \emph{correct} (and \emph{complete}), since all included schemes are by definition.
    The schemes can however differ in whether they provide \emph{public verifiability} and under which conditions \emph{input privacy} and \emph{security} hold.
    An evaluation of all solutions with respect to these properties is provided in \cref{tab:classification} and discussed in more detail in the sections that come after.

    \subsubsection{Assumptions}
    Security and input privacy can be based on different kinds of assumptions.
    Most schemes base their security on (computational) cryptographic assumptions.
    We classify the used assumptions as either \emph{standard} or \emph{non-standard}.
    Standard assumptions are those that have been widely accepted by the cryptographic community, e.g., CDH or DDH; such assumptions have often been accepted for a longer period of time.
    Next to cryptographic assumptions, some schemes, especially those based on \acp{zk-SNARK} may require a \emph{trusted setup}.
    If this trusted setup is broken, malicious parties are able to create false proofs and break security guarantees.

    Alternatively, some solutions rely on \emph{trusted hardware} or \emph{\acp{TTP}} for guaranteeing input privacy and/or security.

    \subsubsection{Efficiency and practicality}
    Practical performance and efficiency also play a big role in which solution is best suitable in a certain setting.
    However, since all solutions use different techniques, and different ways of evaluating the practical performance (if any), it is not feasible to compare the practical performance fairly.
    Instead, we focus on the asymptotic computation and communication complexities of all schemes and summarize these.
    An overview thereof is provided in \cref{sec:asymptotic-complexities-of-vppc-schemes}.

    Lastly, we describe in \cref{tab:classification} whether the original paper includes some form of experimental performance evaluation, whether an implementation is publicly available, or whether there are any other practical aspects that influence the usability of the solution.

    \section{Verifiable Privacy-Preserving Computation}\label{sec:verifiable-privacy-preserving-computation}
    Verifiable protocols for specific applications have been around for several decades.
    To the best of our knowledge, the first use of \emph{public verifiability} was in the context of e-voting~\cite{cohenRobustVerifiableCryptographically1985,schoenmakersSimplePubliclyVerifiable1999,chaumSecretballotReceiptsTrue2004}.
    Public verifiability has also been researched for other applications, such as online auctions~\cite{naorPrivacyPreservingAuctions1999}, and secret sharing~\cite{stadlerPubliclyVerifiableSecret1996, schoenmakersSimplePubliclyVerifiable1999}.

    \subsection{Definitions}\label{subsec:definitions}
    Research into schemes for (public) verifiability of computations has gained traction mostly over the past decade.
    Solutions come from different subfields and under different names, but all have a common ground in that they aim to provide (public) verifiability for generic computations, often over private data.
    Below, we expand on these developments, by starting from \ac{VC} and then providing a general definition for \ac{VPPC} schemes.
    This is followed by a specialized definition for \ac{MPC}-based protocols.
    A more extensive discussion of prior definitions from literature is presented in \cref{sec:other-definitions}.

    \paragraph{Verifiable Computing}
    \Ac{VC} describes the setting where a (computationally constrained) client outsources the evaluation of a function $f$ to an external worker.
    \Iac{VC} scheme should satisfy at least: \emph{correctness}, \emph{security}, and \emph{efficiency} (\cref{def:vc-correctness,def:vc-security,def:vc-efficiency}).
    The basic definition of verifiable computation does not provide \emph{input privacy} with respect to the workers, however it can be additionally defined as \cref{def:vc-privacy}.
    The notion of \iac{VC} scheme (\cref{def:vc-scheme}) is, to our best knowledge, first discussed in~\cite{gennaroNoninteractiveVerifiableComputing2010} in the \emph{designated verifier} setting.
    An extension towards public verifiability was provided in~\cite{gennaroQuadraticSpanPrograms2013} and further generalized in~\cite{parnoPinocchioNearlyPractical2013} to \cref{def:vc-public-verif}.

    \paragraph{Generic \ac{VPPC}}
    As observed in~\cite{viandVerifiableFullyHomomorphic2023}, there is a wide variety of definitions for verifiable \ac{HE} (a more elaborate discussion can be found in \cref{subsec:formal-he}).
    We observe a similar variety for definitions of \ac{DLT}- and \ac{DP}-based \ac{VPPC}.
    To better discuss and compare the different \ac{VPPC} schemes and their properties, we provide a generalizing definition for \ac{VPPC} schemes, based on existing definitions from \ac{VC}~\cite{gennaroQuadraticSpanPrograms2013,gennaroNoninteractiveVerifiableComputing2010,parnoPinocchioNearlyPractical2013} and verifiable \ac{HE}~\cite{viandVerifiableFullyHomomorphic2023}:
    \begin{definition}[VPPC scheme]\label{def:vppc}
    A VPPC scheme $\mathcal{VPPC}$ is a 5-tuple of \ac{ppt} algorithms:
    \begin{itemize}
        \item $\keygen(f, 1^\securityParameter) \rightarrow (\secretKey, \publicKey, \evaluationKey)$: given a function $f$ and security parameter \securityParameter, outputs a secret key \secretKey, public key \publicKey, and evaluation key \evaluationKey;
        \item $\encode(\publicKey/\secretKey, x) \rightarrow (\sigma_x, \tau_x)$: given either \publicKey or \secretKey, and an input $x$, returns an encoded input $\sigma_x$ and attestation $\tau_x$;
        \item $\compute(\evaluationKey, \sigma_x) \rightarrow (\sigma_y, \tau_y)$: given \evaluationKey and $c_x$, returns an encoded output $\sigma_y$ and corresponding attestation $\tau_y$;
        \item $\verify(\sigma_y, \tau_x, \tau_y) \rightarrow \{0,1\}$: given $\sigma_y$, $\tau_x$ and $\tau_y$, returns 1 if $\sigma_y$ is a valid encoding of $f(x)$, and 0 otherwise;
        \item $\decode(\secretKey, \sigma_y) \rightarrow y$: given \secretKey and $\sigma_y$, returns the decoded output $y$.
    \end{itemize}
    \end{definition}

    A scheme $\mathcal{VPPC}$ should satisfy at least, \emph{correctness}, \emph{completeness}, \emph{security}, and \emph{input privacy}.
    We explicitly split the \ac{VC} definition of \emph{correctness} into \emph{correctness} and \emph{completeness} to also allow for schemes with alternative correctness definitions, such as in the case of approximate computations.
    For example, for approximate \ac{HE} or \ac{DP} methods we can define approximate correctness as in~\cite{viandVerifiableFullyHomomorphic2023}.
    Moreover, we note that \ac{DP}-based schemes do not provide strict \textit{input privacy} but a different form of privacy, known as \emph{$\epsilon$-differential privacy}, as defined in, e.g.,~\cite{dworkDifferentialPrivacy2006, katoPreventingManipulationAttack2021}.

    \begin{definition}[Correctness]\label{def:vppc-correctness}
    A scheme $\mathcal{VPPC}$ is correct if for all functions $f$, with $\keygen(f, 1^\securityParameter) \rightarrow (\secretKey, \publicKey, \evaluationKey)$, such that \\ $\forall x \in Domain(f)$, if $\encode(\publicKey/\secretKey, x) \rightarrow (\sigma_x, \tau_x)$ and \\ $\compute(\evaluationKey, \sigma_x) \rightarrow (\sigma_y, \tau_y)$ then $\Pr[\decode(\secretKey, \sigma_y) = f(x)] = 1$.
    \end{definition}

    \begin{definition}[Completeness]\label{def:vppc-completeness}
    A scheme $\mathcal{VPPC}$ is complete if for all functions $f$, with $\keygen(f, 1^\securityParameter) \rightarrow (\secretKey, \publicKey, \evaluationKey)$, such that \\ $\forall x \in Domain(f)$, if $\encode(\publicKey/\secretKey, x) \rightarrow (\sigma_x, \tau_x)$ and \\ $\compute(\evaluationKey, \sigma_x) \rightarrow (\sigma_y, \tau_y)$, then $\Pr[\verify(\sigma_y, \tau_x, \tau_y) = 1] = 1$.
    \end{definition}

    \begin{definition}[Security]\label{def:vppc-security}
    A scheme $\mathcal{VPPC}$ is secure, if for any \ac{ppt} adversary $\mathcal{A} = (\mathcal{A}_1, \mathcal{A}_2)$, such that for all functions $f$, if $\keygen(f, 1^\securityParameter) \rightarrow (\secretKey, \publicKey, \evaluationKey)$, $\mathcal{A}_1^{\mathcal{O}_{\encode},\mathcal{O}_{\decode}}(\publicKey, \evaluationKey) \rightarrow x$, \\ $\encode(\publicKey/\secretKey, x) \rightarrow (\sigma_x, \tau_x)$, $\mathcal{A}_2^{\mathcal{O}_{\encode},\mathcal{O}_{\decode}}(\sigma_x, \tau_x) \rightarrow (\sigma_y, \tau_y)$ then $\Pr[\verify(\sigma_y, \tau_x, \tau_y) = 1 \land \decode(\secretKey, \sigma_y) \neq f(x)] \leq \textsf{negl}(\securityParameter)$.
    \end{definition}

    \begin{definition}[Input privacy]\label{def:vppc-privacy}
    A scheme $\mathcal{VPPC}$ has input privacy, if for any \ac{ppt} adversary $\mathcal{A} = (\mathcal{A}_1, \mathcal{A}_2)$, such that for all functions $f$, if $\keygen(f, 1^\securityParameter) \rightarrow (\secretKey, \publicKey, \evaluationKey)$,\\ $\mathcal{A}_1^{\mathcal{O}_{\encode},\mathcal{O}_{\decode}}(\publicKey, f, 1^\securityParameter) \rightarrow (x_0, x_1)$, \\ $\{0,1\} \xrightarrow{R} b$, $\encode(\publicKey/\secretKey, x_b) \rightarrow (\sigma_b, \tau_b)$ \\ then $2|  \Pr[\mathcal{A}_2^{\mathcal{O}_{\encode}}(\sigma_b,\tau_b) = b] - \frac{1}{2}| \leq \textsf{negl}(\securityParameter)$.
    \end{definition}

    In \cref{def:vppc-security,def:vppc-privacy}, the decoding oracle $\mathcal{\mathcal{O}_{\decode}}(\sigma_y, \tau_x, \tau_y)$ returns $\bot$ if $\verify(\secretKey, \sigma_y, \tau_x, \tau_y) = 0$ and $\decode(\secretKey, \sigma_y)$ otherwise.
    The encoding oracle $\mathcal{O}_{\encode}(\publicKey/\secretKey, x)$ returns $\encode(\publicKey/\secretKey, x)$ for any valid input $x$.

    We say that a scheme has \emph{public verifiability} if $\tau_x$ is public rather than private, and all security definitions still hold, noting that the adversary now also has access to $\tau_b$ in the definition of \emph{input privacy}.

    \paragraph{MPC-based \ac{VPPC}}
    For \ac{MPC}-based \ac{VPPC} schemes we will adopt an alternative definition, that better fits the construction of such schemes, it is provided in \cref{sec:definition-mpc-based-vppc}.
    This definition is inspired by existing definitions for publicly auditable \ac{MPC}~\cite{baumPubliclyAuditableSecure2014,ozdemirExperimentingCollaborativeZkSNARKs2022} and is analogous to our generic definition above.

    \subsection{High-level overview}\label{subsec:high-level-comparison}
    Below, we discuss the typical application settings of the different classes, and additionally compare efficiency at a high level.
    \ac{DLT}- and \ac{DP}-based schemes are applicable to more specialized use cases, whereas \ac{MPC}- and \ac{HE}-based solutions are more general and can often be used in the same situations.

    \paragraph{DLT}
    First, we note that \ac{DLT}-based solutions are best suited for situations with participants that do not have direct communication channels with one another, or for computations with varying groups of participants.
    A downside of \ac{DLT} with respect to the other solutions is the limitation in message sizes and verification times that are imposed by the use of a shared ledger.
    Moreover, the lack of private communication channels often leads to more complicated solutions than are possible when such channels are present.
    In cases where direct communication channels are available, \ac{HE}-, \ac{DP}- and \ac{MPC}-based solutions will generally be more practical and efficient.

    \paragraph{DP}
    As described in \cref{subsec:differential-privacy}, \ac{DP}-based solutions generally make use of a central party to compute the result, irrespective of where the noise is applied.
    Generally, these solutions are significantly more efficient than \ac{MPC}- or \ac{HE}-based approaches, but only provide approximate correctness of the result and provide weaker privacy guarantees for the original inputs.
    In return \ac{DP}-based method can protect against inference and membership attacks, unlike any of the other methods.
    Therefore, in use cases with large amounts of data or clients, complex computations, or risks of inference and membership attacks, \ac{DP}-based methods can outperform alternative constructions.

    \paragraph{MPC vs.\ HE}
    \ac{MPC}-based solutions have a strong focus on computations over distributed data, however can also be applied for verifiable outsourcing.
    In either setting, the computations are performed by a group of workers, of which a fraction is required to be honest to guarantee input privacy.
    The minimum size of this fraction depends on the underlying \ac{MPC} protocol (see \cref{tab:classification}).

    In verifiable outsourcing settings, the use of \iac{HE}-based \ac{VPPC} scheme is often more practical, since computations on \ac{HE} ciphertexts can be executed by a single central server that need not be trusted to guarantee input privacy.
    However, \ac{HE}-based solutions can also be used in settings with distributed data.
    In that case, all data owners use a distributed, or threshold, key generation protocol to obtain a public-private key pair where the private key is shared over all data owners~\cite{asharovMultipartyComputationLow2012, jainThresholdFullyHomomorphic2017}.
    Then, all data owners encrypt their data under the same public key and let the central server perform the actual computation on the received ciphertexts.

    The main difference between \ac{HE}- and \ac{MPC}-based schemes lies in the efficiency of both schemes.
    Generally speaking \ac{MPC}-based schemes require significant amounts of communication, either in multiple online rounds, or in one large offline pre-processing round.
    The computations themselves are often rather simple and can be executed locally by each party.
    Alternatively, \ac{HE}-based schemes can communicate all required ciphertexts in one round and let the computations be executed by a single party.
    Downside of \ac{HE} with respect to \ac{MPC}, is the high computational costs of performing \ac{HE} operations and the large size of \ac{HE} ciphertexts.
    For complicated computations, \ac{MPC} will often be more time-efficient, however the right choice will differ per use case.

    Additionally, we observe that \ac{MPC} schemes have been widely researched for multiple decades and have long reached practical efficiency.
    \ac{HE} schemes are more novel, and have only recently started to reach practical efficiency.
    Being an active area of research, many optimizations for \ac{HE} are yet to be expected.

    \subsection{High-level observations}\label{subsec:overarching-challenges}
    We make a number of high-level observations regarding \ac{VPPC} schemes in general.
    This predominantly concerns topics that are currently underexposed, but are very relevant for their adoption.

    \paragraph{Input data authentication}
    Verifiability of \iac{PPC} guarantees that the (public) output is computed correctly from the secret input.
    In other words, no corrupted party has meddled with the results.
    However, corrupted parties could still produce incorrect or malicious outputs by providing false inputs.

    In most real-world use cases, computation correctness alone will not be sufficient, and input authentication will be required.
    One would protect the entire data pipeline from authenticated input to result, by combining the two.
    Moreover, such solutions can be used to guarantee reproducibility, i.e., it can be guaranteed that computations were verifiably executed on the same data.

    In our analysis we found only one recent solution, \citetalias{duttaComputeVerifyEfficient2022}, that focused on both verifiability and input authentication.

    \paragraph{Reusability}
    In practice, we often run different algorithms on the same data, or the same algorithms on different data.
    Logically, the question can be raised whether reusing parts of the (intermediate) data can improve efficiency, reduce communication, or provide guarantees for reproducibility.
    The solutions we found in our analysis had little to no attention for such reusability.

    However, we do observe a number of non-verifiable solutions for reusable \ac{MPC}~\cite{bartusekReusableTwoRoundMPC2020} or reusable \ac{GC}~\cite{harth-kitzerowCRGCPracticalFramework2022} appear in recent years.
    With increased efficiency and less communication, reusability is especially beneficial in decentralized settings with large amounts of data or many participants.
    Benefits become even more clear, when considering that \acp{VPPC} use costly primitives like \ac{HE} and \ac{ZKP}.

    \paragraph{Post-quantum security}
    In a world where the threat of quantum computers on classical cryptography is increasing rapidly, \ac{PQ} secure solutions are becoming increasingly important.
    This is underscored by, e.g., the NIST standardization for \ac{PQ} primitives~\cite{computersecuritydivisionPostQuantumCryptography2017}, implementation of \ac{PQ} primitives for OpenSSL~\cite{microsoftresearchPostQuantumTLS2024}, and the NCSC migration recommendations~\cite{ncscPreparingQuantumSafeCryptography2020}.
    While sufficiently large quantum computers may seem out of reach now, current predictions expect quantum computers to break non-\ac{PQ} primitives in 30  years~\cite{josephTransitioningOrganizationsPostquantum2022}.
    Especially, in cases where ciphertexts and/or commitments are made publicly available nowadays (as for publicly verifiable \ac{VPPC}), attackers could adopt harvest-now-decrypt-later attacks, to store ciphertexts today and break them in a couple of decades.
    It is thus essential, to design \ac{PQ} secure \ac{VPPC} schemes now, so that we can implement them on time to still provide security levels that increase exponentially in the security parameter, even in the presence of quantum computers.

    Most, recent \ac{FHE} schemes are believed to be \ac{PQ} secure, and information-theoretically secure \ac{MPC} protocols have been around for a long time.
    However, many other primitives used to create \ac{VPPC} schemes are not \ac{PQ} secure.
    More importantly, security against quantum adversaries was not discussed in the majority of the works we saw, even though its relevance increases by the day.

    \paragraph{Comparing practical performances}
    In our analysis, we observed that all works use very different methods to evaluate their asymptotic and/or practical performance (also see \cref{sec:asymptotic-complexities-of-vppc-schemes}).
    A surprisingly large subset of papers does not discuss performance aspects in its entirety.
    We admit it is no small feat to compare different \ac{VPPC} schemes, especially those of different classes.
    However, to make well-informed choices for future research and real-world adoption it is of the utmost importance to be able to fairly compare different solution approaches at least at a high level.

    Making implementations of the majority of the schemes publicly available, would also greatly improve the ability to compare the practical performance of the different solutions, and adopt real-world adoption.
    We specifically mention as an example the MP-SPDZ work~\cite{kellerMPSPDZVersatileFramework2020}, which presented a well maintained framework of many recent \ac{MPC} solutions, making easy comparison of performance and real-world applications available to a wide audience.

    \section{MPC-based VPPC}\label{sec:mpc-based-vppc}

    Solutions that use \ac{MPC} as privacy-preserving computation mechanism for constructing a \ac{VPPC} scheme can be divided in three groups.
    Each group uses a different mechanism for verifiability: succinct \acp{ZKP}, non-succinct \acp{ZKP}, or other.
    The final group consists of schemes that use mechanisms that are different from all the other papers in this class.

    \subsection{Non-succinct ZKP}\label{subsec:non-succinct-zkp-mpc}
    The majority of verifiable \ac{MPC} solutions uses commitments to (their shares) of input, intermediate, and output values in combination with \acp{NIZK} to guarantee security.
    First, we discuss solutions using non-succinct \ac{ZKP}.

    \subsubsection{Description}
    The first set of works (\citetalias{riviniusPubliclyAccountableRobust2022, baumPubliclyAuditableSecure2014, cunninghamCatchingMPCCheaters2016, schoenmakersUniversallyVerifiableMultiparty2015}) uses $\Sigma$-protocols and the \ac{FS} heuristic to obtain \acp{NIZK} from standard assumptions.
    One solution is based on the CDN~\cite{cramerMultipartyComputationThreshold2001} framework.
    The three other works (\citetalias{cunninghamCatchingMPCCheaters2016, baumPubliclyAuditableSecure2014, riviniusPubliclyAccountableRobust2022}) use a more efficient SPDZ-like protocol for their \ac{MPC} computation.
    Two of the SPDZ-like protocols (\citetalias{cunninghamCatchingMPCCheaters2016, baumPubliclyAuditableSecure2014}) additionally rely on \acp{MAC} similar to those used in the original SPDZ protocol.
    We also note that \citetalias{riviniusPubliclyAccountableRobust2022} makes use of \ac{PQ} secure lattice-based commitments.

    A downside of $\Sigma$-protocols is the large proof size and significant effort required on the verifier's side for larger computations.
    Hence, recent works (\citetalias{duttaComputeVerifyEfficient2022, priviledgeprojectRevisionExtendedCore2021}) often apply the more efficient compressed $\Sigma$-protocol theory~\cite{attemaCompressedSProtocolTheory2020}, or $\Sigma$-bullets~\cite{bunzZetherPrivacySmart2020}, while still relying only on standard assumptions.

    Verification of all protocols in this subclass is done by verifying the entire transcript of the \ac{MPC} computation.
    A verifier needs to check the \acp{ZKP} at each step to guarantee that each commitment to a new share is computed correctly.

    \subsubsection{Evaluation}
    Protocols in this subclass provide security and public verifiability even when all parties are corrupted.
    Moreover, all schemes are based on standard cryptographic assumptions and do not have a trusted setup.
    The number of honest parties required to ensure input privacy depends on the underlying \ac{MPC} scheme.

    The protocols relying on standard $\Sigma$-protocols in combination with the \ac{FS} heuristic (\citetalias{riviniusPubliclyAccountableRobust2022, baumPubliclyAuditableSecure2014, cunninghamCatchingMPCCheaters2016, schoenmakersUniversallyVerifiableMultiparty2015}) are generally speaking more costly in verification time and proof size than the schemes relying on compressed $\Sigma$-protocols (\citetalias{duttaComputeVerifyEfficient2022, priviledgeprojectRevisionExtendedCore2021}).

    The efficiency of the \ac{MPC} computations and communication thereof depend mostly on the underlying protocol used.
    The choice of \ac{MPC} scheme also depends on the amount of dishonest parties one accepts regarding the breakage of input privacy.
    For a comparison of these schemes with respect to their privacy guarantees and asymptotic complexities we refer to \cref{tab:classification} and \cref{tab:efficiency}.
    We note that some of the schemes did not report efficiency metrics, or only provided incomparable experimental evaluations.
    An open question is whether we can find a uniform way to compare these schemes with respect to their efficiencies.

    Finally we observe that there is one work, \citetalias{riviniusPubliclyAccountableRobust2022}, using a large number of \ac{PQ} secure primitives.
    Although this does not make the scheme fully \ac{PQ} secure, the authors speculate that it is possible to do so.
    Unfortunately, we found no other works in this class that discussed the risk of quantum computing on the security of their solutions.
    Further research is needed to determine which solutions could be made \ac{PQ} secure.

    \subsection{Succinct ZKP}\label{subsec:succinct-zkp-mpc}
    Next to schemes relying on non-succinct \acp{ZKP}, we observe solutions using succinct \acp{ZKP}.
    These schemes work similarly, but are often more efficient with respect to verification and transcript size.

    \subsubsection{Description}
    All solutions that we observed (\citetalias{veeningenPinocchioBasedAdaptiveZkSNARKs2017, kanjalkarPubliclyAuditableMPCasaService2021, ozdemirExperimentingCollaborativeZkSNARKs2022, schoenmakersTrinocchioPrivacyPreservingOutsourcing2016}), use distributed, often adaptive, \acp{zk-SNARK} to assure security.
    Most of these solutions allow for the use of vector commitments, implying that each party publishes one commitment to all their shares.
    The computation can then be verified by checking the \ac{zk-SNARK} proof given the parties' commitments and the computed output.

    \subsubsection{Evaluation}
    Constructions based on succinct \acp{ZKP} are in many ways similar to those based on non-succinct \acp{ZKP}.
    These proofs guarantee security and public verifiability even when all parties are corrupted.
    The number of honest parties needed to provide input privacy is determined by the underlying \ac{MPC} scheme.

    The difference between succinct and non-succinct \acp{ZKP} lies in the trade-off between security and efficiency.
    Succinct \ac{ZKP}-based solutions have very small proof sizes and verification times.
    Moreover, these solutions work very efficiently with vector commitments, and do not require a verifier to check the entire transcript of the computation, reducing the communication and verification costs even more.
    This all comes at the cost of relying on non-standard cryptographic assumptions and the fact that most \acp{zk-SNARK} require a trusted setup.
    It is an open question who should be involved in this trusted setup to guarantee trust in the correctness proofs.

    The succinct schemes we found in our literature review compare as follows.
    The construction of \citetalias{schoenmakersTrinocchioPrivacyPreservingOutsourcing2016} only guarantees input privacy given an honest majority of computation parties.
    The other works can be used with any \ac{LSSS} scheme, offering the users a choice in the number of parties required to be honest, making this the more flexible choice.

    The main difference between the schemes is the type of distributed \ac{zk-SNARK} that is used.
    \citetalias{veeningenPinocchioBasedAdaptiveZkSNARKs2017} uses a \ac{zk-SNARK} that is non-universal, i.e., a new setup is needed for a new computation, and requires a trusted setup.
    \citetalias{kanjalkarPubliclyAuditableMPCasaService2021} does use a universal \ac{zk-SNARK}, making it possible to perform one universal trusted setup, rather than one per computation.
    Finally, \citetalias{ozdemirExperimentingCollaborativeZkSNARKs2022} allows for the usage of any \ac{zk-SNARK} proof that can be computed in a distributed fashion.
    This allows for flexibility and adoption of novel schemes that may be more efficient or have better security assumptions.
    An open question is whether \ac{zk-SNARK} schemes with a trustless setup or \ac{PQ} security, e.g., Fractal~\cite{chiesaFractalPostquantumTransparent2020}, can be used in this way.

    \subsection{Other}\label{subsec:other-mpc}
    Below, we describe the remaining solutions that did not fit any other category.

    \subsubsection{Description}
    \citetalias{jakobsenFrameworkOutsourcingSecure2014} uses \acp{MAC} to design a scheme where $n$ clients verifiably outsource a computation to $m$ workers.
    However, only clients can verify the computation, and at least one worker needs to be honest to guarantee security and input privacy.

    \citetalias{baumEfficientConstantRoundMPC2020} presents an efficient, constant round, publicly verifiable \ac{MPC} protocol that only makes black-box use of cryptographic primitives.
    The scheme uses BMR-type garbled circuits~\cite{beaverComplexitySecureProtocols1990} and requires black-box access to a statistically secure \ac{OT} scheme and a cryptographic hash function.
    Notably, security and public verifiability only hold when at least one party is honest.

    Alternatively, \citetalias{schabhuserFunctionDependentCommitmentsVerifiable2018} uses pairing-based function-dependent commitments, leading to efficient verification of correctness.
    However, this construction only supports linear functions.

    \citetalias{ramchenUniversallyVerifiableMPC2019} constructs an \ac{MPC} protocol using an \ac{SWHE} scheme supporting one single multiplication.
    Verifiable key switching, using \acp{NIZK}, is used to support computations with an arbitrary number of multiplications.

    Finally, \citetalias{baldimtsiCrowdVerifiableZeroKnowledge2020} presents a solution relying on crowd verifiable \acp{ZKP}.
    Meaning that a prover tries to convince a predefined group of verifiers (\enquote*{crowd}), where each verifier only provides a bounded amount of randomness, thus requiring a mostly honest crowd.

    \subsubsection{Evaluation}
    At first glance, the approach of \citetalias{jakobsenFrameworkOutsourcingSecure2014} seems more (asymptotically) efficient than the other \ac{MPC} schemes.
    But, we observe that this scheme only provides input privacy and security in case of at least one honest client.
    Ergo, it provides similar guarantees as most \ac{MPC} schemes that are secure against a dishonest majority.
    This is not the type of security guarantee we aim for to achieve verifiable, privacy-preserving computations.

    \citetalias{baumEfficientConstantRoundMPC2020} provides similar security guarantees, but also adds public verifiability, which is a feature not found in regular \ac{MPC} schemes.
    It is therefore an interesting approach, but we deem it only useful for the particular use case where one assumes at least one of the computing parties is honest.
    This assumption seems undesirable in most use cases where public verifiability is required.

    Since \citetalias{schabhuserFunctionDependentCommitmentsVerifiable2018} only supports the computation of linear functions, we deem it too limited to be useful in practice.
    It is doubtful that this approach can be adapted to support non-linear function without fundamental changes.

    \citetalias{ramchenUniversallyVerifiableMPC2019} is an interesting approach, in that it provides input privacy given one honest party and public verifiability even when all parties are malicious, i.e., more guarantees cannot be given.
    The use of verifiable key-switching in combination with \iac{SWHE} scheme is more or less similar to that of the SPDZ-like protocols discussed above.
    However, the SPDZ-like protocols move the computationally costly steps to an offline preprocessing phase, whereas they happen in the online phase here.
    The authors of \citetalias{ramchenUniversallyVerifiableMPC2019} do not describe the (asymptotic) complexity of their work.
    Nonetheless, we observe that the above mentioned SPDZ-like protocols have a significantly more efficient online phase, making them more practical.

    The use of crowd verifiable \acp{ZKP} in \citetalias{baldimtsiCrowdVerifiableZeroKnowledge2020} is different from any of the other \ac{MPC} constructions.
    It is suitable only for a very particular use case, where there is a select, predefined group of verifiers that wants to check the correctness of a computation.
    While designated-verifier \acp{ZKP} could be more efficient than publicly verifiable \acp{ZKP}, we do not expect these differences to be significant.
    Moreover, the protocol of \citetalias{baldimtsiCrowdVerifiableZeroKnowledge2020} requires interaction with the verifiers, which is a big downside in most use cases, when compared to other \ac{ZKP} solutions that are non-interactive.
    We therefore expect the other \ac{ZKP} solutions to be more useful in practice.
    Moreover, developments in that direction are applicable to a much wider variety of \ac{VPPC} use cases than is the case for \citetalias{baldimtsiCrowdVerifiableZeroKnowledge2020}.

    \subsection{Comparison}\label{subsec:comparison-mpc}
    We observe that the most promising solutions all rely on \acp{ZKP} to guarantee security and/or public verifiability.
    Specifically, we see two flavors of these constructions: schemes using \acp{zk-SNARK} and schemes using \acp{ZKP} from standard cryptographic assumptions.
    Both types make a different trade-off between efficiency and security.
    Therefore, we do not proclaim a preference towards either solution, as this will depend on use case specific requirements.

    We do note, that within constructions relying on \acp{zk-SNARK}, there is only little focus on dealing with the problematic trusted setup, that is required by the specific \acp{zk-SNARK} that are used.
    However, to obtain true public verifiability, one would require external parties to also be involved in this trusted setup, which is a big inconvenience in practice.

    Second, we observe that constructions that rely on $\Sigma$-protocols in combination with the \ac{FS} heuristic have significantly larger proof size and verification time than those using compressed $\Sigma$-protocols.
    However, the latter category is more novel, leading to a smaller number of works in the category, even though the efficiency is much better, and security guarantees are similar.
    We expect constructions based on compressed $\Sigma$-protocols to become more popular, but further research is needed.

    \subsubsection*{Open challenges}
    \begin{itemize}[topsep=0pt]
        \item \ac{VPPC} from only \ac{PQ} secure primitives, to guarantee security even in the presence of a quantum adversary.
        \item Efficiently combining vector commitments and \acp{ZKP}, for example through commit-and-prove-\acsp{SNARK}~\cite{campanelliLegoSNARKModularDesign2019}.
        \item More efficient verification and proof sizes in constructions based on standard assumptions.
        \item Succinct \ac{ZKP}-based constructions without a trusted setup.
        \item Modular solutions; due to the developments in both \ac{MPC} constructions and especially \acp{ZKP} schemes, modular solutions could more quickly take advantage of novel and improved schemes.
        \item Almost all solutions assume that the provided inputs of the parties are honest, however in practice this need not be the case.
        Further research into dealing with this problem, e.g., by means of authenticated inputs, is needed.
    \end{itemize}

    \section{HE-based VPPC}\label{sec:he-based-vppc}
    \ac{HE}-based schemes can be divided in four groups, based on the mechanism used for verifiability: homomorphic \acp{MAC}, succinct \acp{ZKP}, \acp{TEE}, or other.
    The final group consists of solutions that rely on mechanisms that are different from all other works in this class.

    \subsection{Homomorphic MACs}\label{subsec:homomorphic-macs-he}
    We observe three different approaches for combining \ac{HE} ciphertexts and \acp{MAC}, agreeing with~\cite{viandVerifiableFullyHomomorphic2023}: (1) Encrypt-then-\acs*{MAC}; (2) Encrypt-and-\acs*{MAC}; (3) \acs*{MAC}-then-Encrypt.

    \subsubsection{Description}

    The \textit{Encrypt-then-\acs*{MAC}} method is used by \citetalias{fioreEfficientlyVerifiableComputation2014}.
    In this construction, one first homomorphically encrypts each input and only then computes a homomorphic \ac{MAC} of each ciphertext.
    Therefore, the \ac{MAC} does not have to be hiding.
    The function is then computed by evaluating it on both the input ciphertexts and corresponding \acp{MAC}.
    A correct computation can be verified before decryption, by checking that the resulting \ac{MAC} matches the resulting ciphertext.

    \textit{Encrypt-and-\acs*{MAC}} approaches have a \ac{MAC} and ciphertext that are independent of one another, both are computed directly from the plaintext.
    The \ac{MAC} therefore needs to be both hiding and support the ciphertext operations required for the \ac{HE} scheme.
    Also here, the requested function is computed on both the input \acp{MAC} and the input ciphertexts.
    A correct computation can be verified before decryption, by verifying that the resulting \ac{MAC} matches the resulting ciphertext.

    We found the most occurring approach of the three to be \textit{\acs*{MAC}-then-Encrypt}, where the \ac{MAC} is computed from the plaintext and concatenated to the plaintext before encryption.
    This removes the need for the \ac{MAC} to be either hiding or support complex ciphertext maintenance operations.
    The \ac{MAC} now only needs to support operations in the plaintext domain.
    In this case, the function is computed by executing it only on the input ciphertexts.
    A correct computation, however, can only be verified after decryption, by verifying that the decrypted \ac{MAC} matches the decrypted result.

    \subsubsection{Evaluation}
    We only found one occurrence of the \textit{Encrypt-then-\acs*{MAC}} approach dating from~\citeyear{fioreEfficientlyVerifiableComputation2014}: \citetalias{fioreEfficientlyVerifiableComputation2014}.
    The presented solution is rather limited, in that it only supports the computation of quadratic circuits.
    Due to the lack of other, more recent, works using this approach, it seems doubtful whether this approach can be improved to support general circuits.
    One would need to design a \ac{MAC} scheme that supports all operations of the \ac{HE} scheme used, including ciphertext maintenance operations.
    It is unclear whether this is possible at all.

    Solutions using the \textit{Encrypt-and-\acs*{MAC}} approach suffer from similar issues.
    While the homomorphic \ac{MAC} of \citetalias{liPrivacyPreservingHomomorphicMACs2018} does support any amount of additions and multiplications, it does not support ciphertext maintenance operations, making bootstrapping impossible.
    This severely limits the multiplicative depth of the function to be computed, if one wants the \ac{HE} computations to remain practical.
    To overcome this problem one would need to design a hiding, homomorphic \ac{MAC} supporting these more complex \ac{HE} operations.

    The most promising, and occurring, approach seems to be \textit{\acs*{MAC}-then-Encrypt}, since in this case the homomorphic \ac{MAC} only needs to support plaintext addition and multiplication.
    The first solutions of this type were still limited by technical constraints.

    The first known approach, \citetalias{gennaroFullyHomomorphicMessage2013}, verifies evaluations of boolean circuits.
    However, this approach was no more efficient than running the computation itself.
    An improvement upon this scheme, with more efficient verification of \ac{HE} computations on arithmetic circuits, is given by \citetalias{catalanoPracticalHomomorphicMACs2013}.
    It is however limited to arithmetic circuits of bounded depth.
    \citetalias{chatelVerifiableEncodingsSecure2022}, a more recent \acs*{MAC}-then-Encrypt scheme, does not have any of these limitations and can thus be used to verify \ac{HE} evaluations of any arithmetic circuit.

    Whilst the \textit{\acs*{MAC}-then-Encrypt} approach is the most practical of the three, it should be noted that contrary to the other methods, the \ac{MAC} can now only be checked after decryption and not before.
    Implying that some information could be leaked if the computation is incorrect.
    Finally, we note that none of these schemes are publicly verifiable, due to needing the full secret key to verify the \ac{MAC}.

    \subsection{ZKPs}\label{subsec:zkps-he}
    Only in recent years the first solutions that use \acp{ZKP} for verifiable computations on \ac{HE} ciphertexts have started to appear.

    \subsubsection{Description}
    All solutions that we observed used \acp{zk-SNARK} to achieve verifiability.
    The (distributed) computation is performed directly on the homomorphic ciphertexts.
    The resulting ciphertext is accompanied by \iac{zk-SNARK} proof verifying that it is indeed the result of applying the function to the input ciphertexts.

    The first solutions (\citetalias{fioreBoostingVerifiableComputation2020, boisFlexibleEfficientVerifiable2020}) both require homomorphic hashing to make the ciphertexts fit inside the groups that are used by the specific \acp{zk-SNARK} used.
    To do so, \citetalias{fioreBoostingVerifiableComputation2020} requires very specific setup parameters for the \ac{HE} scheme, leading to inefficient \ac{HE} operations.
    \citetalias{boisFlexibleEfficientVerifiable2020} improves upon this work by not restricting the \ac{HE} parameter space.
    However, both solutions are limited to \enquote*{simpler} \ac{FHE} schemes, i.e., without complex ciphertext maintenance operations, since these are not supported by the homomorphic hashing scheme used.
    \citetalias{ganeshRinocchioSNARKsRing2021} proposes a method that does not rely on homomorphic hashes, but rather uses a \ac{zk-SNARK} that natively operates on ring elements, i.e., \ac{HE} ciphertexts.

    Alternatively, one can also directly encode the \ac{HE} operations into an arithmetic circuit suitable for any circuit-\ac{zk-SNARK}.
    Experimental results for this approach are discussed for \citetalias{viandVerifiableFullyHomomorphic2023}~\cite{viandVerifiableFullyHomomorphic2023}.

    \subsubsection{Evaluation}
    Two of the early solutions (\citetalias{fioreBoostingVerifiableComputation2020, boisFlexibleEfficientVerifiable2020}), suffer from drawbacks that make them highly impractical to use.
    The restriction on the \ac{HE} parameters in \citetalias{fioreBoostingVerifiableComputation2020} makes the \ac{HE} computations too slow to be practical.
    And the lack of support of ciphertext maintenance operations in both schemes, puts an implicit bound on the multiplicative depths of the functions that can be computed.

    The alternative solution, of translating the \ac{HE} operations, including ciphertext maintenance operations directly into an arithmetic circuit (\citetalias{viandVerifiableFullyHomomorphic2023}) makes it possible to support modern, efficient \ac{HE} schemes.
    However, the complexity of \ac{HE} operations, and the fact that \ac{HE} ciphertexts do not naturally fit \ac{zk-SNARK} groups, makes proof generation costly, and impractical for realistic use cases.

    The most promising solution using \acp{ZKP} is \citetalias{ganeshRinocchioSNARKsRing2021}, \iac{zk-SNARK} natively operating on \ac{HE} ciphertexts, thereby drastically reducing prover time for complex computations~\cite{viandVerifiableFullyHomomorphic2023}.
    However, an open question for \citetalias{ganeshRinocchioSNARKsRing2021} is how to do ciphertext maintenance operations.
    This generally necessitates operations not supported by rings or switching between different rings, which is not supported by \citetalias{ganeshRinocchioSNARKsRing2021}.

    An advantage of \ac{zk-SNARK}-based approaches is the fact that the proofs are publicly verifiable, unlike homomorphic \acp{MAC} or \ac{TEE} attestation.
    Moreover, proof sizes are succinct, verification times small, and correctness of the resulting ciphertext can be verified before decryption.
    A downside of \acp{zk-SNARK} is their reliance on non-standard cryptographic assumptions, and the fact that most schemes require a trusted setup.

    \subsection{TEEs}\label{subsec:tees-he}
    \ac{HE} computations can also be verified by executing them inside a \ac{TEE} and using remote attestation to guarantee security.

    \subsubsection{Description}
    \citetalias{natarajanCHEXMIXCombiningHomomorphic2021} presents such a construction for \ac{FHE}-inside-\ac{TEE}.
    Clients send their encrypted data to \iac{TEE} in the cloud, and only need to attest that the \ac{TEE} performs exactly the desired computation.
    This verification can take place before decrypting the results, allowing all parties to check correctness beforehand.

    \subsubsection{Evaluation}
    The biggest advantage of using \iac{TEE} is computation time, since \acp{TEE} natively support most operations needed for \ac{HE} computations.
    Even though such operations will be slower than on a regular CPU, they are faster than computing, e.g., a zero-knowledge proof.
    Optimizing \ac{FHE} computations for \acp{TEE} can lead to notable performance improvements~\cite{viandVerifiableFullyHomomorphic2023}, making the need for research into this topic clear.

    The use of \iac{TEE} guarantees input privacy and correctness of the computation, given that one trusts the specialized hardware being used.
    It is however unclear how to achieve public verifiability, i.e., how attestation can be (efficiently) performed by external parties, or long after the computation has been executed.

    \subsection{Other}\label{subsec:other-he}
    Below, we describe the remaining solutions that did not fit in any of the above categories.

    \subsubsection{Description}
    \citetalias{louVFHEVerifiableFully2023} uses an approach for verifiable \ac{FHE}, called \enquote*{blind hashing}, where the hash of the plaintext is appended to the plaintext before encryption, i.e., this is similar to \ac{MAC}-then-Encrypt.
    The hash is subsequently verified by also computing the hash of the decrypted result and ensuring it equals the hash that was included in the ciphertext.

    \citetalias{gennaroNoninteractiveVerifiableComputing2010} uses Yao's \ac{GC} for one-time verifiable computations between a client and server.
    Subsequently, they propose to add \ac{FHE} on top to transform this one-time approach into a reusable one, allowing multiple evaluations of the same function on different input data.
    It should be noted that a different computation needs a new setup, and computing the setup is very inefficient.

    \subsubsection{Evaluation}
    \citetalias{louVFHEVerifiableFully2023} is a preliminary publication that lacks a security proof/reasoning, and only explains how to perform a single matrix multiplication.
    It is unclear how this approach extends to other operations, such as needed for ciphertext maintenance and more complex circuits.
    We deem an actual \ac{MAC}-then-Encrypt approach more viable than this approach purely based on hashing.

    Being published in~\citeyear{gennaroNoninteractiveVerifiableComputing2010}, makes \citetalias{gennaroNoninteractiveVerifiableComputing2010} the oldest verifiable \ac{HE} approach we found.
    Whilst this scheme is efficient (\cref{def:vc-efficiency}), the fact that each different computation requires the computation of a rather inefficient setup, makes it impractical for real-world usage.
    Our literature search showed no later work using a similar approach.
    All together this makes us conclude that this approach is not as viable as other \ac{HE}-based solutions.

    \subsection{Comparison}\label{subsec:comparison-he}
    Out of the different methods for \ac{HE}-based \ac{VPPC} schemes we observe three predominant categories, either using homomorphic \acp{MAC}, \acp{zk-SNARK}, or \acp{TEE}.

    Of the homomorphic \ac{MAC} approaches, the \acs*{MAC}-then-Encrypt approach seems to be the most promising, due to it putting the least requirements on the homomorphic \ac{MAC} used.
    A downside compared to the other methods is that one does need to decrypt the ciphertext before verifying the \ac{MAC}, it is an open question how we can resolve this issue.
    Another problem with \ac{MAC}-based approaches, is the fact that one needs to know the secret key to allow for verification of the \ac{MAC}, making public verifiability not directly possible.
    We expect it to be possible to solve this issue using other cryptographic techniques, such as \acp{ZKP} or \ac{MPC}.
    But, further research into this topic is needed.

    \acp{zk-SNARK}-based approaches do offer public verifiability.
    However, current solutions suffer from efficiency issues, making them impractical for larger computations; especially when ciphertext maintenance operations are required.
    The overhead caused by proof generation is much larger than that of homomorphic \acp{MAC} or \acp{TEE} attestation.
    However, when public verifiability is a requirement, \acp{zk-SNARK} are currently the only solution.
    We do expect that further improvements in \acp{zk-SNARK} allow for more efficient proof generation, making this method catch up regarding efficiency.

    Another downside of \acp{zk-SNARK} with respect to \acp{MAC} is the fact that \acp{zk-SNARK} often require a trusted setup and are based on non-standard assumptions.
    Moreover, we speculate that current (non-succinct) \acp{ZKP} schemes, based on standard assumptions, require too large proof sizes and computational efforts to be feasible.
    Further research into efficient versions of those schemes with respect to \ac{HE} operations could make such solutions possible.

    \ac{TEE}-based solutions seem to be the most practical of the three, especially with optimizations in place.
    A downside of \ac{TEE}-based solutions, with respect to the other solutions, is the requirement to trust specific hardware.
    This trust may not always be present in practice.
    Another downside of \ac{TEE}-based approach with respect to \acp{zk-SNARK} is the lack of public verifiability.

    All in all, if public verifiability is required, currently \acp{zk-SNARK} seem to be the best solution.
    However, further research into the other directions is expected to lead to more efficient alternatives.
    When public verifiability is not a requirement, the main trade-off is between trust in specific hardware and efficiency.
    This trade-off will be use case dependent and cannot be made in general.

    \subsubsection*{Open challenges}
    \begin{itemize}[topsep=0pt]
        \item Solutions relying on \acp{ZKP} from standard cryptographic assumptions and without a trusted setup.
        \item More efficient \acp{ZKP} that natively support ring operations; with a specific focus on ciphertext maintenance operations and ring switching operations.
        \item Public verifiability using homomorphic \acp{MAC} or \acp{TEE}.
        \item Realizing \acs*{MAC}-then-Encrypt constructions for which the \ac{MAC} can be verified before decrypting the results.
        \item Optimization of \ac{HE} operation inside \acp{TEE}.
    \end{itemize}

    \section{DLT-based VPPC}\label{sec:dlt-based-vppc}
    \Ac{VPPC} schemes that use \ac{DLT} as basis for their distributed computations can be divided in three groups, based on the mechanism used for verifiability.

    \subsection{Succinct ZKPs}\label{subsec:succinct-zkps}
    While most \ac{DLT} applications using succinct \acp{ZKP} focus purely on financial transactions, e.g., Zerocash~\cite{ben-sassonZerocashDecentralizedAnonymous2014}, we also identified works that focus on privacy and verifiability for smart contracts.

    \subsubsection{Description}
    \citetalias{kosbaHawkBlockchainModel2016} is a smart contract system that can be initialized over any decentralized cryptocurrency.
    It has a basic protocol for transferring coins to another party anonymously in the same fashion as Zerocash.
    However, it also allows for combining a coin transfer with programmable logic, in which the function to be computed is provided as a smart contract.
    To achieve this, all users participating do not directly spend a private coin, but commit both a coin and their private function input to a smart contract, along with \iac{zk-SNARK} proof of correct construction.
    Each user also includes an opening for the function input encrypted under the public key of a trusted manager.
    The actual computation of the function takes place off-chain by this trusted manager, who first opens the commitments to all inputs.
    The manager then computes the function and publishes the output, along with \iac{zk-SNARK} proof and a spending distribution of the committed coins.
    Thereby completing both the transaction and computation, whilst keeping the inputs private, i.e., only known by the trusted manager.

    \citetalias{boweZEXEEnablingDecentralized2020} uses a different construction, called \ac{DPC}, that does not rely on a trusted manager for input privacy, and keeps the function itself private too.
    In \iac{DPC} each blockchain entry is a record containing input data, a birth predicate, and a death predicate.
    The birth predicate defines under which conditions this record was constructed.
    The death predicate defines under which conditions a record can be consumed.
    Any user of the blockchain system can consume any records for which the death predicates can be satisfied and use these to create new records, with new payload and predicates, i.e., perform a state transition.
    Since death predicates can put conditions on all details of the newly created records and on the users that can perform a state transition, we can use this system to model a smart contract.
    \citetalias{boweZEXEEnablingDecentralized2020} guarantees anonymity of the payload, or input data, by only adding commitments to the records on-chain.
    Any valid transaction that consumes and creates records, has to add a \ac{zk-SNARK} proof attesting to the correct creation of the new records, and the fact that the user also knows a valid proof for the predicates of each consumed record.

    \subsubsection{Evaluation}
    By using \acp{zk-SNARK}, \citetalias{kosbaHawkBlockchainModel2016} offers public verifiability of correctness of the executed computations, given that the used \ac{zk-SNARK} is secure, irrespective of the behavior of the trusted manager.
    Moreover, since \acp{zk-SNARK} have very small proof sizes and verification times, they are suitable to use in \iac{DLT} environment.
    Privacy of function inputs is guaranteed, but is dependent upon the trusted manager not revealing any of these inputs, which is unrealistic in practice.

    \citetalias{boweZEXEEnablingDecentralized2020} guarantees input privacy without such a trusted manager.
    Moreover, it also adds function privacy to any other observer on the blockchain.
    Only the party consuming a record needs to know the functions, or predicates, used to create this record.
    A downside is that, where in \citetalias{kosbaHawkBlockchainModel2016} a complete computation is performed at once, in \citetalias{boweZEXEEnablingDecentralized2020} a computation is performed per party, given the (intermediate) state.
    This leads to longer computation times.
    Moreover, \citetalias{boweZEXEEnablingDecentralized2020} does not by default keep record data private from the party consuming the record.
    One would still have to rely on \ac{HE}- or \ac{MPC}-style computations to achieve this.

    Public verification of the correctness of the executed computations is very efficient.
    Next to that, the actual function can be computed locally.
    However, most computation time will be consumed by proof generation.
    In the case of \citetalias{boweZEXEEnablingDecentralized2020}, waiting on verification of previous blocks with input records for the next step might lead to a large amount of latency on top of this.

    Finally, we note that \acp{zk-SNARK} are based on non-standard cryptographic assumptions, which may be undesirable in practice.
    Next to this, most \acp{zk-SNARK} require a trusted setup, which, if broken, could be used to create false proofs.

    \subsection{Non-succinct ZKPs}\label{subsec:non-succinct-zkps-dlt}
    We also found one solution based on non-succinct \acp{ZKP}, making a different trade-off between security and efficiency.

    \subsubsection{Description}
    \citetalias{bunzZetherPrivacySmart2020} is a privacy-preserving payment system for smart contract platforms using $\Sigma$-bullets.
    $\Sigma$-bullets combine the optimizations of bulletproofs~\cite{bunzBulletproofsShortProofs2018} with classic $\Sigma$-protocol theory, to create more efficient $\Sigma$-style proofs.
    Where non-private cryptocurrencies require access to the transaction details to verify each transaction, \citetalias{bunzZetherPrivacySmart2020} only adds commitments to the transaction details on the public ledger.
    These commitments are accompanied by $\Sigma$-bullets, that attest to exactly the predicates that are normally checked in verification.
    Rather than checking these predicates directly, any verifier can now check correctness of the provided proof to obtain the same guarantees.

    \subsubsection{Evaluation}
    The advantage of using $\Sigma$-bullets rather than \acp{zk-SNARK} is that they only rely on standard security assumptions and do not require a trusted setup, leading to stronger privacy and security guarantees, whilst still providing public verifiability.
    This comes at the cost of more expensive verification and larger proof sizes.
    Both of these increase with the size of the computation that is executed.
    \citetalias{bunzZetherPrivacySmart2020} only offers private coin transfer, and does not support generic computations.

    While $\Sigma$-bullets can be used for generic computations~\cite{attemaCompressedSProtocolTheory2020}, this would lead to verification times and proof sizes that are likely too large to be used in \iac{DLT} setting.
    It is doubtful whether non-succinct \acp{ZKP} are suitable for generic computations in \iac{DLT} setting.

    \subsection{TEEs}\label{subsec:tees-dlt}
    \citetalias{chengEkidenPlatformConfidentialityPreserving2019} is a blockchain solution where smart contracts over private data are executed off-chain.

    \subsubsection{Description}
    In \citetalias{chengEkidenPlatformConfidentialityPreserving2019}, specialized compute nodes with \acp{TEE} execute the smart contract and compute the results.
    The consensus nodes use remote attestation to verify these results and update the smart contract results on-chain accordingly.

    \subsubsection{Evaluation}
    Rather than relying on expensive cryptographic machinery or trusted parties, \citetalias{chengEkidenPlatformConfidentialityPreserving2019} uses trusted hardware to guarantee correctness, whilst maintaining privacy.
    Whilst \acp{TEE} are slower than computations on regular CPUs, they are multiple orders of magnitudes faster than generating zero-knowledge proofs~\cite{viandVerifiableFullyHomomorphic2023}.
    This comes however at the cost of relying upon the privacy and security guarantees of the trusted hardware.
    If the hardware is compromised or faulty, privacy and security could be broken as a whole.
    To what extent one can and should trust \ac{TEE} hardware is an open question and might require the technology to be around for longer to gain more trust.

    Another open question is that of public verifiability, \acp{TEE} have very different remote attestation mechanisms.
    It is unclear whether one can achieve public verifiability, or even verify correctness long after the computation has taken place.
    Moreover, since not all parties may have the means or access to perform remote attestation, \citetalias{chengEkidenPlatformConfidentialityPreserving2019} puts trust in a select group of nodes to perform verification.
    This may not be desirable in all cases.

    \subsection{Comparison}\label{subsec:comparison-dlt}

    \ac{DLT} applications often require very small verification time and small message sizes in order to make block verification practical, and keep the size of the ledger manageable.
    Due to the larger verification time and proof sizes of non-succinct \acp{ZKP} approaches, we do not expect such solutions to be feasible in \iac{DLT} setting for generic computations.
    When considering \acp{ZKP}, \acp{zk-SNARK} seem the logical choice in \iac{DLT} setting.
    Especially the \ac{DPC} approach as described by \citetalias{boweZEXEEnablingDecentralized2020} seems very promising.
    Not only does it provide input privacy and public verifiability, but also guarantees function privacy, something that has not been observed in any of the other classes.
    Open questions, however, exist regarding how to improve the efficiency of composable \acp{zk-SNARK} and how to remove trusted setups.
    Moreover, \ac{DPC}-based approaches do not hide the input data (or function to be computed), from the other parties involved in the computation.
    This could be solved using, e.g., an \ac{MPC}-based computation.
    However, using \ac{MPC} directly, without \ac{DLT}, might be more efficient in that case.
    There should be clear, additional benefits of using \ac{DLT} before choosing it over other approaches.

    An alternative approach to \iac{ZKP}-based approach, is to perform computations using \iac{TEE}.
    This is more efficient than using \acp{ZKP}, but does require the user to have trust in and access to specific trusted hardware.
    It is doubtful whether this is practical in \iac{DLT} setting, mostly due to the fact that \ac{TEE} hardware is often only available centrally, thereby defeating the purpose of \ac{DLT}.

    Both promising approaches require trust in other means that standard cryptographic assumptions.
    We do not expect this to be circumventable in the near future, due to the current requirements on message sizes and verification times in \ac{DLT} settings.

    \subsubsection*{Open challenges}
    \begin{itemize}[topsep=0pt]
        \item More efficient composable \ac{zk-SNARK} constructions.
        \item Schemes using \acp{ZKP} without trusted setup.
        \item Efficient solutions to keep data and/or functions private from other computation parties.
    \end{itemize}

    \section{DP-based VPPC}\label{sec:dp-based-vppc}
    All \ac{VPPC} schemes based on \ac{DP} make use of \acp{ZKP} in some form.

    \subsubsection{Description}
    In \citetalias{narayanVerifiableDifferentialPrivacy2015}, a \emph{trusted curator} holds all data of the individual parties, and publishes a public commitment to the dataset.
    A vetted analyst can study the data and determine a \ac{DP}-private query it wants to evaluate.
    The curator confirms that the query is indeed differentially private and subsequently generates the corresponding keys for the \ac{ZKP} generation.
    Next, the analyst executes the query and publishes the \ac{DP} result along with a publicly verifiable \ac{ZKP}, which can be verified by any external party.

    \citetalias{tsaloliDifferentialPrivacyMeets2023} uses a similar setting, but additionally ensures that the input data is kept private from the analyst.
    However, the authors do not provide sufficient details to determine which concrete building blocks could be used to realize this scheme.

    \citetalias{biswasVerifiableDifferentialPrivacy2023} presents a scheme for verifiable \ac{DP} in the trusted curator model, and offers an alternative in which the clients secret share their data over multiple servers that together \emph{emulate} the trusted curator using \ac{MPC}, thereby increasing the privacy guarantees.
    Rather than relying on \ac{NIZK} proofs as in the previous two constructions,~\citetalias{biswasVerifiableDifferentialPrivacy2023} rely on interactive $\Sigma$-protocols to convince a public verifier of the validity of the results.
    If this public verifier is a trusted party, this scheme additionally obtains public verifiability by releasing the transcripts of an interaction.

    As an alternative to the trusted curator setting, \citetalias{movsowitzdavidowPrivacyPreservingTransactionsVerifiable2023} presents a framework for verifiable \ac{LDP} in \iac{DLT} setting.
    They show how a client holding certain private attributes, that are committed to on a public blockchain, can reveal a verifiably \ac{DP} version of an attribute to an analyst.
    Client and analyst first perform a protocol to generate true randomness together.
    At a later point in time, the client uses this randomness to verifiable randomize a selected attribute.
    Public verifiability is guaranteed using \acp{zk-SNARK}.
    While verification of the subsequent computations by the analyst are not discussed, we expect that this could be done using similar techniques.

    Alternatively, \citetalias{katoPreventingManipulationAttack2021} relies on $\Sigma$-protocols and cryptographic randomized response~\cite{ambainisCryptographicRandomizedResponse2004} techniques, to ensure that clients provide honestly randomized values to the analyst.
    They describe how to verifiable execute three state-of-the-art \ac{LDP} protocols on discrete, private inputs.

    \subsubsection{Evaluation}
    The field of \ac{DP}-based \ac{VPPC} schemes is still rather novel and there seems to be much potential for improvements and alternative constructions, to provide stronger privacy and security guarantees or be applicable on more diverse settings.
    We observe that the \ac{LDP} methods (\citetalias{movsowitzdavidowPrivacyPreservingTransactionsVerifiable2023,katoPreventingManipulationAttack2021}) provide stronger privacy guarantees for the clients' data.
    However, both works only present ways to protect the privacy for binary or a small number of discrete values, using a specific type of randomized response.
    Extensions to other settings, or alternative perturbation methods would allow for applicability in a much wider range of settings.
    Moreover, these methods focus purely on verifiable randomization of the input data, but do not provide similar guarantees for the function that is evaluated on that data.

    In the trusted curator model (\citetalias{narayanVerifiableDifferentialPrivacy2015,tsaloliDifferentialPrivacyMeets2023,biswasVerifiableDifferentialPrivacy2023}), we do observe verification of the function evaluation.
    However, this comes at the cost of a trusted curator, who gets access to the private input data.
    There are possibilities for providing even stronger privacy guarantees in this setting, using \ac{MPC} or \ac{HE}, thereby improving upon \citetalias{biswasVerifiableDifferentialPrivacy2023} which already distributes the curator role.

    Finally, we observe that most of these methods require interaction between client and curator, which can become a burden in the case of many clients.
    In order to make these methods practical for real world applications, a reduction in the amount of communication would be needed.

    \subsubsection*{Open challenges}
    \begin{itemize}[topsep=0pt]
        \item Verifiable \ac{LDP} for real-valued input data.
        \item Stronger privacy guarantees in the trusted curator model.
        \item Reducing communication between clients and curator.
    \end{itemize}

    \begin{table}
    \footnotesize
    \centering
    \caption{High-level comparison of solutions paradigms.}
    \label{tab:summary_paradigms}
    \begin{threeparttable}
        \begin{tabular}{lllllll}
            \toprule
            Solution paradigm    & \rot{Practical}                 & \rot{Public verifiability}  & \rot{Communication cost}       & \rot{Prover cost}                & \rot{Verifier cost}          & \rot{Assumptions$^\dagger$}   \\ \midrule
            MPC+Non-succinct ZKP & \cellcolor[HTML]{67FD9A}Yes     & \cellcolor[HTML]{67FD9A}Yes & \cellcolor[HTML]{FD6864}High   & \cellcolor[HTML]{FFCC67}Medium   & \cellcolor[HTML]{FD6864}High & \cellcolor[HTML]{67FD9A}S      \\
            MPC+Succinct ZKP     & \cellcolor[HTML]{67FD9A}Yes     & \cellcolor[HTML]{67FD9A}Yes & \cellcolor[HTML]{67FD9A}Low    & \cellcolor[HTML]{FFCC67}Medium   & \cellcolor[HTML]{67FD9A}Low  & \cellcolor[HTML]{FFCC67}NS+TS  \\
            HE+MAC               & \cellcolor[HTML]{FFCC67}Partial & \cellcolor[HTML]{FD6864}No  & \cellcolor[HTML]{67FD9A}Low    & \cellcolor[HTML]{67FD9A}Low      & \cellcolor[HTML]{67FD9A}Low  & \cellcolor[HTML]{FFCC67}S      \\
            HE+ZKP               & \cellcolor[HTML]{FFCC67}Partial & \cellcolor[HTML]{67FD9A}Yes & \cellcolor[HTML]{67FD9A}Low    & \cellcolor[HTML]{FD6864}High     & \cellcolor[HTML]{67FD9A}Low  & \cellcolor[HTML]{FFCC67}NS+TS  \\
            HE+TEE               & \cellcolor[HTML]{67FD9A}Yes     & \cellcolor[HTML]{FD6864}No  & \cellcolor[HTML]{67FD9A}Low    & \cellcolor[HTML]{67FD9A}Low      & \cellcolor[HTML]{67FD9A}Low  & \cellcolor[HTML]{FFCC67}TH     \\
            DLT+Non-succinct ZKP & \cellcolor[HTML]{FD6864}No      & \cellcolor[HTML]{67FD9A}Yes & \cellcolor[HTML]{FD6864}High   & \cellcolor[HTML]{FFCC67}Medium   & \cellcolor[HTML]{FD6864}High & \cellcolor[HTML]{67FD9A}S      \\
            DLT+Succinct ZKP     & \cellcolor[HTML]{67FD9A}Yes     & \cellcolor[HTML]{67FD9A}Yes & \cellcolor[HTML]{67FD9A}Low    & \cellcolor[HTML]{FFCC67}Medium   & \cellcolor[HTML]{67FD9A}Low  & \cellcolor[HTML]{FFCC67}NS+TH  \\
            DLT+TEE              & \cellcolor[HTML]{FD6864}No      & \cellcolor[HTML]{FD6864}No  & \cellcolor[HTML]{67FD9A}Low    & \cellcolor[HTML]{67FD9A}Low      & \cellcolor[HTML]{67FD9A}Low  & \cellcolor[HTML]{FFCC67}TH     \\
            DP+ZKP               & \cellcolor[HTML]{67FD9A}Yes     & \cellcolor[HTML]{67FD9A}Yes & \cellcolor[HTML]{FFCC67}Medium & \cellcolor[HTML]{67FD9A}Low      & \cellcolor[HTML]{67FD9A}Low  & \cellcolor[HTML]{FFCC67}Varies \\ \bottomrule
        \end{tabular}
        \begin{tablenotes}
            \item[$\dagger$] S = Standard; NS = Non-Standard; TS = Trusted Setup; TH = Trusted Hardware
        \end{tablenotes}
    \end{threeparttable}
\end{table}

    \begin{table}
    \footnotesize
    \centering
    \captionsetup{justification=centering}
    \caption{Summary of open challenges per solution paradigm.}
    \label{tab:summary_challenges}
    \begin{threeparttable}
        \begin{tabular}{ll}
            \toprule
            \textbf{Challenge}                         & \textbf{Solution paradigm(s)}  \\ \midrule
            \Acl{PQ} security                          & All$^\ddagger$                 \\
            Modularity                                 & All$^\ddagger$                 \\
            Dealing with dishonest inputs              & All$^\ddagger$                 \\
            More efficient proof generation            & Any$^\dagger$+ZKP              \\
            More efficient verification                & Any$^\dagger$+Non-succinct ZKP \\
            Smaller communication size                 & Any$^\dagger$+Non-succinct ZKP \\
            Dealing with or removing trusted setup     & Any$^\dagger$+Succinct ZKP     \\
            Public verifiability                       & DLT+TEE; HE+TEE; HE+\ac{MAC}   \\
            Combine (vector) commitments with \ac{ZKP} & MPC+ZKP                        \\
            Support ciphertext maintenance operations  & HE+\ac{MAC}                    \\
            Keep data private from curator             & DP+ZKP                         \\
            Non-discrete input data                    & DP+ZKP                         \\ \bottomrule
        \end{tabular}
        \begin{tablenotes}[para]
            \item[$\dagger$] Any = \{MPC, HE, DLT, DP\}
            \item[$\ddagger$] All = Any + any verifiability paradigm
        \end{tablenotes}
    \end{threeparttable}
\end{table}

    \section{Conclusion}\label{sec:conclusion}
    We presented a systematic overview of \ac{VPPC} schemes, applicable to settings with distributed data and identified four main classes: \ac{MPC}-, \ac{HE}-, \ac{DLT}, and \ac{DP}-based.
    Each class is best suited to different scenarios, depending on the available resources.
    Next, we analyzed the solutions in each class, by dividing the classes based on the used verifiability paradigm and identified the most pressing open challenges.
    A high-level summary is depicted in \cref{tab:summary_paradigms,tab:summary_challenges}.

    \ac{DP}-based \ac{VPPC} schemes are generally the most efficient, but often offer a weaker form of privacy.
    Moreover, these schemes are rather novel and require more research to be suitable for practical use cases.
    For \ac{DLT}-based approaches the use of succinct \acp{ZKP} for verifiability seemed the most promising approach, given the constraints on verification time and proof size.
    \ac{MPC}-based solutions predominantly use \acp{ZKP} for verifiability.
    Those using succinct \acp{ZKP} are significantly more efficient than those based on non-succinct \acp{ZKP}.
    However, this comes at the cost of non-standard security assumptions and trusted setups.
    Finally, for \ac{HE}-based approaches, constructions using \acp{TEE} or homomorphic \acs*{MAC}-then-Encrypt seem most promising, but still suffer from practical limitations.

\begin{acks}
    The authors thank Mohammed Alghazwi, Vincent Dunning, and Berry Schoenmakers for their feedback on initial versions of this paper.
\end{acks}

    \bibliographystyle{ACM-Reference-Format}
    \bibliography{main}

    \appendix

    \section{Solutions adjacent to VPPC}\label{sec:adjacent-solutions}
    In this section we briefly discuss a number of interesting solution types that are adjacent or complementary to \ac{VPPC} schemes.
    They are not \ac{VPPC} schemes themselves, either because they are only suitable for specific computations, or because they can only be used for very specific tasks.
    This list is by no means exhaustive, but is meant to provide a high-level overview of related solution approaches that could be combined with the schemes in \cref{tab:classification} or provide a good alternative in specific use cases.

    The \emph{\ac{FL}} paradigm describes how \ac{ML} models can be trained collaboratively by a group of data owners~\cite{konecnyFederatedOptimizationDistributed2016}, without sharing their respective private datasets.
    Generally speaking, local models are trained at each data owner, such that private data need not be shared.
    On top of that, model updates are shared in a clever way, in order to construct one global model that is similar to one trained on the combined data.
    Since private data need not be distributed or shared with other parties, a certain degree of input privacy is provided.
    We observe however, that this input privacy is not absolute and shared model updates as well as the global model might leak significant information about the original input data~\cite{liuThreatsAttacksDefenses2022}.
    Recently, we observe more attention for the verification of \ac{FL} schemes, concerning either verification of the individual clients or of the central orchestrator.
    An overview of verifiable federated learning approaches is provided in, e.g.,~\cite{zhangVerifiableFederatedLearning2022}.

    \emph{\Ac{PSI}} is another example of a privacy-preserving solution for a specific problem, namely that of determining the intersection over a number of private sets~\cite{moralesPrivateSetIntersection2023,vosSoKCollusionresistantMultiparty2023}.
    Each set is controlled by a different party, and the other parties should learn nothing more than the eventual intersection, which may also be secret shared in some cases.
    Oftentimes \ac{PSI} approaches are only secure in the semi-honest model, but some solutions also provide security in the malicious model (see \cref{sec:threat-models}).
    Additionally, we observe a number of protocols that focus on verifiable, delegated \ac{PSI}, i.e., when the intersection is executed by an external party, and should be verifiable to other parties, e.g.,~\cite{abadiVDPSIVerifiableDelegated2017, teradaImprovedVerifiableDelegated2018}.
    For a more extensive overview, we refer to~\cite{moralesPrivateSetIntersection2023}.

    Next to the above specific solutions, we also note a number of interesting primitives that could be used to extend \ac{VPPC} scheme with data authenticity properties or stronger privacy guarantees.
    Specifically we mention the use of \emph{\ac{PIR}} protocols~\cite{ostrovskySurveySingleDatabasePrivate2007} or \emph{\ac{ORAM}} primitives~\cite{changObliviousRAMDissection2016}.
    Both of these methods allow users or algorithms to access data, in a database or in memory without revealing which data they accessed.
    We note that there already exist schemes that implement \ac{ORAM} using \iac{TEE}~\cite{hoangMOSEPracticalMultiUser2020}.
    In relation to this, we note the existence of \emph{\acp{PoR}}, which allow clients to outsource their data to an external server while ensuring the integrity, soundness and retrievability of said data, e.g.,~\cite{anthoineDynamicProofsRetrievability2021,patersonMultiproverProofRetrievability2018}.
    The auditing capabilities of these \acp{PoR} could allow for strong input authenticity guarantees in combination with \ac{VPPC} schemes.

    Overall, these solutions can be used to make \ac{VPPC} schemes more suitable for practical use cases.
    Combining the techniques used above with any of the \ac{VPPC} schemes would offer an all-in-one solution for treating the entire pipeline of \aclp{VPPC} rather than treating verifiable computation as a standalone problem.

    \section{Additional background on MPC constructions}\label{sec:background-mpc}
    Below, we summarize the most frequently used constructions to create secure \ac{MPC} protocols: secret sharing, garbled circuits, and \ac{PHE}.
    We briefly explain the most important terms and reference notable constructions.
    There also exist so called \emph{mixed protocols}, that combine different \ac{MPC} schemes in an efficient manner to make the best use of the benefits that each scheme provides.
    We do not explain mixed protocols here, but rather refer to a well known example of a mixed protocol framework, that uses efficient conversions between arithmetic secret sharing, boolean secret sharing, and garbled circuits: ABY~\cite{demmlerABYFrameworkEfficient2015}.

    \subsubsection*{Secret Sharing}
    In secret sharing-based \ac{MPC}, private data is split over multiple parties.
    The data is split in such a way that only predefined sets of parties can reconstruct the original data, while other (sub)sets cannot even deduce partial information.
    Most secret sharing schemes divide a secret \secret among $n$ parties in such a way that only subsets with $t$ or more parties can reconstruct \secret.
    Such a scheme is called a \threshold{t}{n} scheme.

    \emph{Additive secret sharing} is one of the most intuitive examples of a secret sharing scheme.
    Given prime $q$ and finite field $\mathbb{Z}_q$, a secret $\secret \in \mathbb{Z}_q$ is shared among $n$ parties $P_1, \ldots, P_n$ by sending a random number $s_i \in \mathbb{Z}_q$ to each party $P_i$ such that $\sum_i s_i = s \pmod{q}$.
    Clearly, the secret can only be reconstructed by using the individual shares of all parties, while any strict subset of parties cannot even deduce partial information on \secret, i.e., it is an example of an \threshold{n}{n} scheme.
    It is also a \emph{\ac{LSSS}}, meaning that any linear operation performed on individual shares is applied to the secret when reconstructed.

    Multiplication of shares often requires the use of online secure multiplication protocol, which is fairly efficient for honest majority situations.
    In case of a dishonest majority, \iac{LSSS} shares can be multiplied efficiently using so called Beaver's multiplication triplets~\cite{beaverEfficientMultipartyProtocols1992}.
    These triplets are input/function-independent and can thus be generated in a so called \emph{offline} preprocessing phase.
    This significantly reduces the cost of the more expensive \emph{online} phase.

    Examples of popular secret sharing-based schemes are: Shamir's secret sharing~\cite{shamirHowShareSecret1979}, SPDZ~\cite{damgardMultipartyComputationSomewhat2012}, and MASCOT~\cite{kellerMASCOTFasterMalicious2016}.
    For an up-to-date overview on existing schemes with implementations we refer to, e.g., the MP-SPDZ library~\cite{kellerMPSPDZVersatileFramework2020}.

    \subsubsection*{Garbled Circuits}
    Yao's \acp{GC}~\cite{yaoHowGenerateExchange1986} are a way to enable two distrusting parties to securely evaluate a function on their private input data.
    It requires that the underlying function is expressed as a boolean circuit consisting of 1-out-2-in gates.

    One party, the \emph{garbler}, generates a garbled version of the entire circuit, by garbling the individual gates.
    In the original implementation, a gate is garbled by assigning a unique, random label to the possible values (true, false) of each input wire, and doubly encrypting the possible output values, under the corresponding input labels.
    The garbler randomly permutes the encrypted outputs and shares these with the \emph{evaluator} accompanied by the random labels corresponding to the private inputs of the garbler.
    The evaluator then participates in an \ac{OT} protocol~\cite{ishaiExtendingObliviousTransfers2003} with the garbler to obtain the labels corresponding to their private inputs.

    Having received both input labels, the evaluator can correctly decrypt exactly one of the garbled outputs, thereby obtaining the true output bit(s).
    When a gate output is used as input to another gate, the true output will not be a bit, but rather a new random label that is used in the decryption of this other gate.

    An alternative construction to Yao's \acp{GC} is the BMR framework~\cite{beaverComplexitySecureProtocols1990}.
    Implementations of both and other frameworks can be found in, e.g., MP-SPDZ~\cite{kellerMPSPDZVersatileFramework2020}, ABY~\cite{demmlerABYFrameworkEfficient2015}, or $\text{ABY}^3$~\cite{mohasselABY3MixedProtocol2018}.

    We note that there are also extensions of garbled circuits to more than two parties~\cite{ben-efraimMultipartyGarblingArithmetic2018}.

    \paragraph{PHE-based \ac{MPC}}
    \Ac{MPC} can also be based on \ac{PHE}.
    These solutions often consist of interactive protocols, making use of the homomorphic properties of \ac{PHE} ciphertexts to reduce the communication.
    Most \ac{PHE}-based \ac{MPC} protocols consist of custom protocols for specific computations, e.g., division~\cite{dahlSecureTwoPartyInteger2012} or comparison~\cite{veugenCorrectionImprovingDGK2018}.
    However, \ac{MPC} frameworks for generic computations also exist, one such example based on threshold Paillier is the CDN framework~\cite{cramerMultipartyComputationThreshold2001}.

    \section{Threat models}\label{sec:threat-models}
    We informally describe the three most common security models regarding the possible behavior of the parties that have been corrupted by the adversary.
    For a more detailed treatment of the topic, we refer the reader to e.g.,~\cite{lindellSecureMultipartyComputation2020}:

    \begin{itemize}
        \item \textit{Semi-honest adversaries:} In this scenario, also known as the \emph{honest-but-curious} or \emph{passive} model, corrupted parties do not deviate from the selected protocol.
        The adversary does have access to all data received and owned by the corrupted parties, and tries to deduce as much as possible from this.
        \item \textit{Malicious adversaries:} In the malicious or \emph{active} setting, the adversary can make corrupted parties deviate from the protocol in any way.
        A protocol is secure if it can prevent any malicious adversary from breaking security or privacy.
        When a protocol not only prevents cheating but also allows for determining who cheated, the protocol is said to support \emph{cheater detection}, e.g.,~\cite{spiniCheaterDetectionSPDZ2016}.
        \item \textit{Covert adversaries:} Corrupted parties show the same behavior as in the malicious model.
        However, honest parties need only detect cheating parties with a given probability.
        In practice, it is assumed that the risk of getting caught and receiving a financial or reputation penalty is sufficient to deter an adversary.
        The notion of covert security with public verifiability was introduced in~\cite{asharovCallingOutCheaters2012}.
        With \emph{public verifiability}, not only can honest parties detect cheating with some probability, they can also construct a publicly verifiable certificate proving that a certain party has cheated, without revealing information on the private data of any honest party.
    \end{itemize}

    \section{Definitions for MPC-based VPPC}\label{sec:definition-mpc-based-vppc}
    \begin{definition}[MPC-based \ac{VPPC}]\label{def:mpc-vppc}
    An MPC-based \ac{VPPC} scheme $\mathcal{VPPC}$ for $N$ parties is a 3-tuple of \ac{ppt} algorithms:
    \begin{itemize}
        \item $\setup(f, 1^\securityParameter) \rightarrow \publicParameters$: given a function $f: X^N \rightarrow Y$ and security parameter \securityParameter, returns public parameters \publicParameters;
        \item $\Pi(\publicParameters, \vec{x}) \rightarrow (y, \tau_y, \verificationKey)$: Given \publicParameters and inputs $\vec{x}$, all parties together execute a protocol to compute the output value $y$, a corresponding attestation $\tau_y$ and a (possibly private) \verificationKey;
        \item $\verify(\publicParameters, y, \tau_y, \verificationKey) \rightarrow \{0,1\}$: Given \publicParameters, $y$, $\tau_y$, and \verificationKey, returns 1 if $f(\vec{x}) = y$, and 0 otherwise.
    \end{itemize}
    \end{definition}

    \Iac{MPC}-based scheme $\mathcal{VPPC}$ should satisfy at least: \emph{correctness}, \emph{security} (against $t \leq N$ malicious parties), and \emph{input privacy}.

    \begin{definition}[Correctness]\label{def:mpc-vppc-correctness}
    A scheme $\mathcal{VPPC}$ is correct if $\Pi$ (restricted to its output $y$) is a correct \ac{MPC}-protocol for evaluating $f$; and if for all functions $f$, with $\setup(f, 1^\securityParameter) \rightarrow (\publicParameters)$, \\ such that $\forall x \in Domain(f)$, if $\Pi(\publicParameters, \vec{x}) \rightarrow (y, \tau_y, \verificationKey)$ \\ then $\Pr[\verify(\publicParameters, y, \tau_y, \verificationKey) = 1] = 1$.
    \end{definition}

    \begin{definition}[Security]\label{def:mpc-vppc-security}
    An MPC-based scheme $\mathcal{VPPC}$ for $N$ parties is secure (against $t\leq N$ malicious parties), if for any \ac{ppt} adversary (controlling up to $t$ parties), such that for all functions $f$, if $\setup(f, 1^\securityParameter) \rightarrow (\publicParameters)$, $\Pi(\publicParameters, \vec{x}) \rightarrow (y, \tau_y, \verificationKey)$ \\ then $\Pr[\verify(\sigma_y, \tau_x, \tau_y) = 1 \land y \neq f(x)] \leq \textsf{negl}(\securityParameter)$.
    \end{definition}

    \begin{definition}[Input Privacy]\label{def:mpc-vppc-privacy}
    An MPC-based scheme $\mathcal{VPPC}$ for $N$ parties has input privacy (against $t\leq N$ malicious parties), if $\Pi$ is a secure \ac{MPC} protocol for evaluating $f$ in the presence of $t$ malicious parties.
    \end{definition}

    We say that a scheme has \emph{public verifiability} if $\verificationKey$ is public rather than private, and all of the above definitions still hold.

    \section{Asymptotic complexities of VPPC schemes}\label{sec:asymptotic-complexities-of-vppc-schemes}
    We provide an overview of the asymptotic complexities of each scheme in \cref{tab:efficiency}.
    In case the authors did not mention the complexity of their solution in the article, and we could not find it in related works, we marked it as \enquote*{N/A}.

    Specifically, we describe the computation and communication complexity of the verifiable computation.
    Whenever possible, we split the costs between those used to create the function result $y$ and those used for creating the verification attestation $\tau_y$.
    If a solution makes use of a generic building block, rather than specifying the exact scheme used, we describe the type of building block used, e.g., \ac{LSSS}, \ac{HE}, or \ac{zk-SNARK}.
    Since verification is executed apart from the computation in the case of public verifiability, we list its complexity in a separate column.

    Finally, we list the asymptotic number of commitments used by each solution, if they are used and if their complexity is not yet included in the other columns.

    \begin{table*}
    \footnotesize
    \centering
    \caption{Asymptotic complexities of each \ac{VPPC} scheme.$^\dagger$}
    \label{tab:efficiency}
    \begin{threeparttable}
        \begin{tabular}{lllllll} \toprule
        Name                                                                     & \multicolumn{2}{l}{Computation}                                                           & \multicolumn{2}{l}{Communication}                                                       & Verification                                                              & \# Comm.   \\
                                                                                 & $y$              & $\tau_y$                                                               & $y$             & $\tau_y$                                                              &                                                                           &            \\ \midrule
        \citetalias{veeningenPinocchioBasedAdaptiveZkSNARKs2017}                 & LSSS             & $O(m \cdot \log(m))$ $\mathbb{F}$, $O(m)$ $\mathbb{G}$                 & LSSS            & $O(n)$ $\mathbb{G}$                                                   & $O(n)$ pairings                                                           & $O(n)$     \\
        \citetalias{kanjalkarPubliclyAuditableMPCasaService2021}                 & LSSS             & \tnote{c}                                                              & LSSS            & $O(1)$ $\mathbb{G}$, $O(1)$ $\mathbb{F}$                              & $O(n)$ pairings, $O(\log(|C|))$ $\mathbb{F}$                              & $O(n)$     \\
        \citetalias{ozdemirExperimentingCollaborativeZkSNARKs2022}               & LSSS             & zk-SNARK                                                               & LSSS            & zk-SNARK                                                              & zk-SNARK                                                                  & $O(l)$     \\
        \citetalias{jakobsenFrameworkOutsourcingSecure2014}                      & LSSS             & $O(1)$ sec.\ add./ open./tag ver.\tnote{b}                             & LSSS            & $O(l)$                                                                & $O(1)$                                                                    &            \\
        \citetalias{catalanoPracticalHomomorphicMACs2013}                        & $O(C)$           & $O(a d + m d \log(d))$                                                 & HE              & $O(a d + m d \log(d))$                                                & $O(|C| + d)$                                                              &            \\
        \citetalias{fioreEfficientlyVerifiableComputation2014}                   & HE               & $O(d_r |C|)$                                                           & HE              & $O(1)$                                                                & N/A                                                                       &            \\
        \citetalias{chatelVerifiableEncodingsSecure2022}                         & HE               & REP: $O(\securityParameter)$: PE: $O(m)$\tnote{d}                      & HE              & REP: $O(\securityParameter)$: PE: $O(m)$\tnote{d}                     & REP: $O(\securityParameter)$: PE: $O(m)$\tnote{d}                         &            \\
        \citetalias{fioreBoostingVerifiableComputation2020}                      & HE               & $O(d_r l + |C|)$                                                       & HE              & $O(1)$ $\mathbb{G}$ + $O(1)$ $\mathbb{F}$                             & $O(1)$ pairings, $O(1)$ exp.\                                             &            \\
        \citetalias{ganeshRinocchioSNARKsRing2021}                               & HE               & N/A                                                                    & HE              & $O(1)$                                                                & N/A                                                                       &            \\
        \citetalias{viandVerifiableFullyHomomorphic2023}                         & HE               & circuit zk-SNARK                                                       & HE              & circuit zk-SNARK                                                      & circuit zk-SNARK                                                          &            \\
        \citetalias{kosbaHawkBlockchainModel2016}                                & Regular          & circuit zk-SNARK                                                       & Regular         & circuit zk-SNARK                                                      & circuit zk-SNARK                                                          &            \\
        \citetalias{boweZEXEEnablingDecentralized2020}                           & Regular          & zk-SNARK over circuit SNARK\tnote{e}                                   & Regular         & zk-SNARK\tnote{e}                                                     & zk-SNARK\tnote{e}                                                         &            \\
        \citetalias{bunzZetherPrivacySmart2020}                                  & Regular          & $O(|C|)$                                                               & Regular         & $O(\log(|C|))$                                                        & $O(|C|)$                                                                  &            \\
        \citetalias{narayanVerifiableDifferentialPrivacy2015}                    & Regular          & circuit zk-SNARK                                                       & Regular         & circuit zk-SNARK                                                      & circuit zk-SNARK                                                          &            \\
        \citetalias{movsowitzdavidowPrivacyPreservingTransactionsVerifiable2023} & Regular          & circuit zk-SNARK                                                       & Regular         & $O(n)$ + circuit zk-SNARK                                             & circuit zk-SNARK                                                          &            \\
        \citetalias{tsaloliDifferentialPrivacyMeets2023}                         & Regular          & N/A                                                                    & Regular         & N/A                                                                   & N/A                                                                       &            \\
        \citetalias{cuvelierVerifiableMultipartyComputation2016}                 & Regular          & $O(l) + 1$ $\Sigma$-protocols\tnote{a}                                 & Regular         & $O(l) + 1$ $\Sigma$-protocols\tnote{a}                                & $O(l) + 1$ $\Sigma$-protocols\tnote{a}                                    & $O(l)$     \\ \midrule\morecmidrules\midrule
        \citetalias{baumPubliclyAuditableSecure2014}                             & \multicolumn{2}{l}{$O(n |C|)$ $\mathbb{F}$, $O(n \log(\securityParameter))$ $\mathbb{G}$} & \multicolumn{2}{l}{$O(n |C| + n)$ $\mathbb{F}$ + $O(n)$ $\mathbb{G}$}                   & verify transcript                                                         &            \\
        \citetalias{schoenmakersUniversallyVerifiableMultiparty2015}             & \multicolumn{2}{l}{N/A}                                                                   & \multicolumn{2}{l}{N/A}                                                                 & N/A                                                                       &            \\
        \citetalias{cunninghamCatchingMPCCheaters2016}                           & \multicolumn{2}{l}{$O(n m)$}                                                              & \multicolumn{2}{l}{$O(n m)$ offline; $O(n m)$ online\tnote{f}}                          & verify transcript                                                         &            \\
        \citetalias{priviledgeprojectRevisionExtendedCore2021}                   & \multicolumn{2}{l}{$O(m+n)$}                                                              & \multicolumn{2}{l}{$O(\log(m+l))$}                                                      & $O(l)$                                                                    &            \\
        \citetalias{riviniusPubliclyAccountableRobust2022}                       & \multicolumn{2}{l}{$O(n m)$}                                                              & \multicolumn{2}{l}{$O(n^2 m)$ offline; $O(n m)$ online\tnote{f}}                        & verify transcript                                                         &            \\
        \citetalias{duttaComputeVerifyEfficient2022}                             & \multicolumn{2}{l}{$O(|\text{input}| + n)$ exp., $O(1)$ pairings}                         & \multicolumn{2}{l}{$O(n\log(|\text{input}|)\log(\securityParameter))$}                  & verify transcript                                                         &            \\
        \citetalias{schoenmakersTrinocchioPrivacyPreservingOutsourcing2016}      & \multicolumn{2}{l}{N/A}                                                                   & \multicolumn{2}{l}{N/A}                                                                 & N/A                                                                       &            \\
        \citetalias{schabhuserFunctionDependentCommitmentsVerifiable2018}        & \multicolumn{2}{l}{N/A}                                                                   & \multicolumn{2}{l}{N/A}                                                                 & N/A                                                                       &            \\
        \citetalias{ramchenUniversallyVerifiableMPC2019}                         & \multicolumn{2}{l}{N/A}                                                                   & \multicolumn{2}{l}{N/A}                                                                 & N/A                                                                       &            \\
        \citetalias{baldimtsiCrowdVerifiableZeroKnowledge2020}                   & \multicolumn{2}{l}{$O(\securityParameter (n + \# \text{servers}) m)$}                     & \multicolumn{2}{l}{$O(\securityParameter (n + \# \text{servers}) m)$}                   & N/A                                                                       &            \\
        \citetalias{baumEfficientConstantRoundMPC2020}                           & \multicolumn{2}{l}{$O(n^2 m)$}                                                            & \multicolumn{2}{l}{$O(n^2 m)$ offline; $O(n^2 m)$ online\tnote{f}}                      & verify transcript                                                         &            \\
        \citetalias{gennaroFullyHomomorphicMessage2013}                          & \multicolumn{2}{l}{N/A}                                                                   & \multicolumn{2}{l}{$O(\securityParameter)$}                                             & N/A                                                                       &            \\
        \citetalias{liPrivacyPreservingHomomorphicMACs2018}                      & \multicolumn{2}{l}{N/A}                                                                   & \multicolumn{2}{l}{N/A}                                                                 & $O(1)$ exp., $O(1)$ pairings                                              &            \\
        \citetalias{boisFlexibleEfficientVerifiable2020}                         & \multicolumn{2}{l}{$O((l + d^2)d_r + \securityParameter |C|)$}                            & \multicolumn{2}{l}{$O(d^2 d_r + \securityParameter \cdot \textsf{depth}(C) \cdot |C|)$} & $O((l + d^2)d_r + \securityParameter \cdot  \textsf{depth}(C) \cdot |C|)$ &            \\
        \citetalias{natarajanCHEXMIXCombiningHomomorphic2021}                    & \multicolumn{2}{l}{N/A}                                                                   & \multicolumn{2}{l}{N/A}                                                                 & N/A                                                                       &            \\
        \citetalias{louVFHEVerifiableFully2023}                                  & \multicolumn{2}{l}{N/A}                                                                   & \multicolumn{2}{l}{N/A}                                                                 & N/A                                                                       &            \\
        \citetalias{gennaroNoninteractiveVerifiableComputing2010}                & \multicolumn{2}{l}{$O((|C| + |\text{input}|) \cdot \textsf{poly}(\securityParameter))$}   & \multicolumn{2}{l}{$O(|C| \cdot \textsf{poly}(\securityParameter))$}                    & $O(|\text{output}| \cdot \textsf{poly}(\securityParameter))$              &            \\
        \citetalias{chengEkidenPlatformConfidentialityPreserving2019}            & \multicolumn{2}{l}{TEE}                                                                   & \multicolumn{2}{l}{N/A}                                                                 & TEE attestation                                                           &            \\
        \citetalias{biswasVerifiableDifferentialPrivacy2023}                     & \multicolumn{2}{l}{N/A}                                                                   & \multicolumn{2}{l}{N/A}                                                                 & N/A                                                                       &            \\
        \citetalias{katoPreventingManipulationAttack2021}                        & \multicolumn{2}{l}{differs per computation}                                               & \multicolumn{2}{l}{differs}                                                             & differs                                                                   &            \\ \bottomrule
        \end{tabular}
        \begin{tablenotes}
            \item[$\dagger$] $n$= \# parties; $m$ = \# mult.; $a$ = \# add.; $l$ = \# inputs; $C$ = circuit; $d$ = $degree(C)$; $d_r$ = $degree(g)$ for $\mathbb{Z}_q[X]/(g)$; $\mathbb{F}$ = field ops.; $\mathbb{G}$ =  group ops.
            \item[a] One $\Sigma$-protocol to prove correctness of $C$, and $O(l)$ $\Sigma$-protocols to prove correctness of each commitment.
            \item[b] Tag verification takes $O(l)$ when detecting cheater.
            \item[c] $O(1)$ $\text{v-MSM}(3|C|)$ + $\text{v-MSM}(m)$ $\mathbb{G}$, $O(m + |C| \log |C|)$ $\mathbb{F}$; $\text{v-MSM}(m)$: variable base multi-scalar exponentiation (can be computed in $\frac{m}{\log(m)}$ using Pippenger's exponentiation algorithm~\cite{bernsteinPippengerExponentiationAlgorithm2002}).
            \item[d] \citetalias{chatelVerifiableEncodingsSecure2022} presents two constructions: a (1) replication encoding-based (REP); and a (2) polynomial encoding-based (PE) authenticator.
            \item[e] Construction uses one \enquote*{inner} SNARK proof to verify the computation, and one \enquote*{outer} zk-SNARK to verify correctness of the inner SNARK, communication and verification only requires the outer zk-SNARK\@.
            \item[f] Broadcast messages only.
        \end{tablenotes}
    \end{threeparttable}
\end{table*}

    \section{Prior definitions}\label{sec:other-definitions}
    Below, we repeat definitions related to \acl{VC} and \acl{VPPC} schemes from prior works and show how they are related to one another.
    This gives a more extensive overview of how the definitions are formalized, and how these formalizations evolved over time.

    We first give the definitions for \emph{Verifiable Computing} or \emph{Verifiable Outsourcing} in \cref{subsec:formal-vc}.
    Subsequently, we discuss the evolution of definitions for publicly verifiable \ac{MPC} in \cref{subsec:formal-he}.
    Finally, we discuss how existing works have formalized verifiable \ac{HE} in \cref{subsec:formal-he}.

    \subsection{Prior definitions for Verifiable Computing}\label{subsec:formal-vc}
    \begin{definition}[Verifiable Computation Scheme~\cite{gennaroNoninteractiveVerifiableComputing2010,gennaroQuadraticSpanPrograms2013}]\label{def:vc-scheme}
   \Iac{VC} scheme $\mathcal{VC}$ is a 4-tuple of \ac{ppt} algorithms:
    \begin{itemize}
        \item $\keygen(f, 1^\securityParameter) \rightarrow (\secretKey, \publicKey)$: Given security parameter \securityParameter and function $f$, this generates a secret key \secretKey and public key \publicKey;
        \item $\probgen(\secretKey, x) \rightarrow (\sigma_x, \verificationKey)$: Given the secret key $\secretKey$, and input $x$, the client computes an encoding of the input $\sigma_x$ and a private verification key \verificationKey;
        \item $\compute(\publicKey, \sigma_x) \rightarrow \sigma_y$: Given the evaluation key \publicKey and $\sigma_x$, the worker computes the encoded output $\sigma_y$;
        \item $\verify(\secretKey, \verificationKey, \sigma_y) \rightarrow y \cup \bot$: Given the private verification key \verificationKey and $\sigma_y$, the client computes the function output $y$, or obtains $\bot$ when the worker's response is invalid.
    \end{itemize}
    \end{definition}

    \Iac{VC} scheme should satisfy at least three properties: \emph{correctness}, \emph{security}, and \emph{efficiency}.
    Correctness guarantees that the output of an honest worker will satisfy the verification function.
    \begin{definition}[Correctness~\cite{gennaroNoninteractiveVerifiableComputing2010}]\label{def:vc-correctness}
    A verifiable computation scheme $\mathcal{VC}$ is correct if for any function $f$, with $\keygen(f, \securityParameter) \rightarrow (\secretKey, \publicKey)$, such that $\forall x \in Domain(f)$, if $\probgen(\secretKey, x) \rightarrow (\sigma_x, \verificationKey)$ and \\ $\compute(\publicKey, \sigma_x) \rightarrow \sigma_y$, then $\verify(\secretKey, \verificationKey, \sigma_y) = f(x)$.
    \end{definition}

    Security guarantees that the worker cannot make the verification function accept an incorrect result (except for with negligible probability).
    \begin{definition}[Security~\cite{gennaroNoninteractiveVerifiableComputing2010,gennaroQuadraticSpanPrograms2013}]\label{def:vc-security}
    Given a verifiable computation scheme $\mathcal{VC}$ and a \ac{ppt} adversary $\mathcal{A} = (\mathcal{A}_1, \mathcal{A}_2)$, and consider the following experiment for a given function $f$:
    \begin{align*}
        &\textbf{\textsf{Exp}}_{\mathcal{A}}^\text{Verif}\left[\mathcal{VC}, f, \securityParameter\right] \\
        &\qquad\keygen(f, \securityParameter) \rightarrow (\secretKey, \publicKey) \\
        &\qquad\text{For } i=1,\ldots,\ell = \textsf{poly}(\securityParameter) \\
        &\qquad\qquad\mathcal{A}_1(\publicKey) \rightarrow (x_i) \\
        &\qquad\qquad\probgen(\publicKey, x_i) \rightarrow (\sigma_{x_i}, \verificationKey) \\
        &\qquad\qquad\mathcal{A}_2(\publicKey, \sigma_{x_i}) \rightarrow \sigma_{y_i} \\
        &\qquad\qquad\verify(\secretKey, \verificationKey, \sigma_{y_i}) \rightarrow y_i \\
        &\qquad\text{If $\exists i \in \left[1,\ell\right]$ such that $y_i \neq \bot \land y_i \neq F(x_i)$, output \enquote*{1}, else \enquote*{0}.}
    \end{align*}
    A verifiable computation scheme $\mathcal{VC}$ is secure for a class of functions $\mathcal{F}$, if for every $f \in \mathcal{F}$, and every \ac{ppt} adversary $\mathcal{A} = (\mathcal{A}_1, \mathcal{A}_2)$: $\Pr[\textbf{\textsf{Exp}}_{\mathcal{A}}^\text{Verif}\left[\mathcal{VC}, f, \securityParameter\right] = 1] \leq \textsf{negl}(\securityParameter)$.
    \end{definition}

    Efficiency, also known as outsourceability, guarantees that the time spend on encoding the input and performing the verification is smaller than the time to evaluate $f$ directly on $x$.
    \begin{definition}[Efficiency~\cite{gennaroNoninteractiveVerifiableComputing2010}]\label{def:vc-efficiency}
    A verifiable computation scheme $\mathcal{VC}$ is efficient, if for any function $f$ and $\keygen(f, 1^\securityParameter) \rightarrow (\secretKey, \publicKey)$, such that for every input $x$ and any output encoding $\sigma_y$, the time required to evaluate $\probgen(\secretKey, x) \rightarrow (\sigma_x, \verificationKey)$ plus the time required to evaluate $\verify(\secretKey, \verificationKey, \sigma_y)$ is $o(T)$, where $T$ is the fastest known time required to evaluate $f(x)$.
    \end{definition}

    The basic definition of verifiable computation does not provide input privacy with respect to the workers.
    However, we can define input privacy for \ac{VC} schemes using an indistinguishability argument:
    \begin{definition}[Input Privacy~\cite{gennaroNoninteractiveVerifiableComputing2010}]\label{def:vc-privacy}
    Given a verifiable computation scheme $\mathcal{VC}$ and a \ac{ppt} adversary $\mathcal{A} = (\mathcal{A}_1, \mathcal{A}_2)$, and consider the following experiment for a given function $f$, where the adversary has access to an oracle $\textsf{Pub\probgen}_{\secretKey}(x)$ that calls $\probgen(\secretKey,x) \rightarrow (\sigma_x, \verificationKey)$ and only returns $\sigma_x$:
    \begin{align*}
        &\textbf{\textsf{Exp}}_A^\text{Priv}\left[\mathcal{VC}, f, \securityParameter\right] \\
        &\qquad\keygen(f, \securityParameter) \rightarrow (\secretKey, \publicKey) \\
        &\qquad\qquad\mathcal{A}^{\textsf{PubProbGen}_{\secretKey}(\cdot)}_1(\publicKey) \rightarrow (x_0, x_1) \\
        &\qquad\qquad\left\{0,1\right\} \xrightarrow{R} b \\
        &\qquad\qquad\probgen(\secretKey, x_b) \rightarrow (\sigma_b, \verificationKey) \\
        &\qquad\qquad\mathcal{A}^{\textsf{PubProbGen}_{\secretKey}(\cdot)}_2(\publicKey, \sigma_b) \rightarrow \hat{b} \\
        &\qquad\text{If $\hat{b} = b$, output \enquote*{1}, else \enquote*{0}.}
    \end{align*}
    A verifiable computation scheme $\mathcal{VC}$ has input privacy for a class of functions $\mathcal{F}$, if for every $f \in \mathcal{F}$, and every \ac{ppt} adversary $\mathcal{A} = (\mathcal{A}_1, \mathcal{A}_2)$: $2|\Pr[\textbf{\textsf{Exp}}_A^\text{Priv}\left[\mathcal{VC}, f, \securityParameter\right] = 1]-\frac{1}{2}| \leq \textsf{negl}(\securityParameter)$.
    \end{definition}

    This definition however, is applicable only in the \emph{designated verifier} setting, i.e., the client is also the verifier.
    An extension towards public verifiability was provided in~\cite{gennaroQuadraticSpanPrograms2013}, where the difference with the designated verifier version is that the secret key is empty, and the verification key can be made public, thereby allowing any party to verify the result.
    Moreover, in the definitions of security and privacy, the adversary now also gets access to verification key.
    In~\cite{parnoPinocchioNearlyPractical2013}, the definition for \emph{Public Verifiable Computation Scheme} was further generalized.
    This lead to a more compact definition of verifiable computation:

    \begin{definition}[Publicly Verifiable Computation Scheme~\cite{parnoPinocchioNearlyPractical2013}]\label{def:vc-public-verif}
    A publicly verifiable computing scheme $\mathcal{VC}$ is a 3-tuple of \ac{ppt} algorithms:
    \begin{itemize}
        \item $\keygen(f, 1^\securityParameter) \rightarrow (\evaluationKey, \verificationKey)$: Given security parameter \securityParameter and function $f$, th generates a public evaluation key \evaluationKey, and a public verification key \verificationKey;
        \item $\compute(\evaluationKey, x) \rightarrow (y, \pi)$: Given the public evaluation key \evaluationKey and input $x$, the worker computes the function output $y$ and a proof $\pi$ attesting to the correctness thereof;
        \item $\verify(\verificationKey, x, y, \zkproof)$: Given the public verification key \verificationKey, input $x$, output $y$ and proof $\zkproof$, this returns 1 if $y = f(x)$ and 0 otherwise.
    \end{itemize}
    \end{definition}

    The definition for \emph{correctness}, \emph{security}, and \emph{efficiency} are highly similar in~\cite{gennaroNoninteractiveVerifiableComputing2010}, we therefore provide a summarized version here:
    \begin{itemize}
        \item \textbf{Correctness}: For any function $f$, with $\keygen(f, 1^\securityParameter) \rightarrow (\evaluationKey, \verificationKey)$, such that $\forall x \in Domain(f)$, if $\compute(\evaluationKey, x) \rightarrow (y, \pi)$,\\ then $\verify(\verificationKey, x, y, \zkproof) = 1$.
        \item \textbf{Security}: For any function $f$ and any \ac{ppt} adversary $\mathcal{A}$:
        \begin{multline*}
            \Pr\big[\mathcal{A}(\evaluationKey, \verificationKey) \rightarrow (x, y, \zkproof): f(x) \neq y \\ \land \verify(\verificationKey, x, y, \zkproof) = 1 \big] \leq \textsf{negl}(\securityParameter).
        \end{multline*}
        \item \textbf{Efficiency}: For any function $f$, input value $x$, output value $y$ and proof $\zkproof$, the time required to run $\verify(\verificationKey, x, y, \zkproof)$ is smaller than the time needed to regularly evaluate $f(x)$.
    \end{itemize}

    \subsection{Prior definitions for publicly verifiable MPC}\label{subsec:formal-mpc}
    Verifiability for arbitrary computations in the \ac{MPC} framework was first studied in~\cite{hooghDesignLargeScale2012}, and more concretely defined in~\cite{baumPubliclyAuditableSecure2014} who refer to it as \emph{public auditability} (see \cref{def:mpc-public-verif}).
    We observe that this definition coincides with the \emph{security} definition in the public verifier setting as used in \ac{VC}.
    Next to this we observe that, unlike for \ac{VC}, \ac{MPC} computations always take input privacy into account.

    Clearly, \ac{MPC} protocols should also provide \emph{correctness}.
    In the semi-honest threat model (see \cref{sec:threat-models}), the parties honestly follow the protocol, therefore one only has to proof that the protocol itself delivers the correct result.
    This coincides with the \ac{VC} definition of correctness.

    However, in the malicious model (see \cref{sec:threat-models}), the parties may deviate from the protocol.
    In this case one has to show, that all honest parties will still obtain the correct result (or the computation is aborted, in the case of security-with-abort), in the presence of corrupted parties.
    This corresponds with the \ac{VC} definition of security in the designated verifier setting.

    Finally, we observe that, unlike for \ac{VC}, efficiency is not given as a particular requirement for a (publicly verifiable) \ac{MPC} scheme, since \ac{MPC} focusses on computation over distributed data from multiple parties, rather than outsourcing a computation of a single client.
    This, of course, does not mean that efficiency is not considered as a factor here, however it is not required (and often not possible) that the \ac{MPC} protocol is more efficient that a straightforward evaluation of the function it implements.

    Additionally, we observe that in the field of \acp{GC}, or \emph{garbling schemes}, there exists a property known as \emph{authenticity} which is formally described in~\cite{bellareFoundationsGarbledCircuits2012}.
    Authenticity captures whether an adversary could break the correctness of the protocol and its definition is similar to the security definition of~\cite{gennaroNoninteractiveVerifiableComputing2010} (\cref{def:vc-security}) on which it is based.
    For a more extensive overview on \ac{GC}-based \ac{MPC}, and other properties specific to such schemes, we refer the reader to~\cite{bellareFoundationsGarbledCircuits2012}.

    We include here the, slightly rephrased, definition of \emph{public verifiability} from~\cite{baumPubliclyAuditableSecure2014}:

    \begin{definition}[Public verifiability~\cite{baumPubliclyAuditableSecure2014}]\label{def:mpc-public-verif}
    Let $\mathcal{C}$ be a circuit, and let $x_1, \ldots, x_m$ be inputs to $\mathcal{C}$, $y$ be a potential output of $\mathcal{C}$ and $\tau$ be a protocol transcript for the evaluation of the circuit $\mathcal{C}$.
    We say that an $\ac{MPC}$ protocol satisfies \emph{public verifiability} if the following holds: an auditor $\mathcal{P}^A$ with input $\tau$ accepts $y$ (with overwhelming probability) if the circuit $\mathcal{C}$ on input $x_1, \ldots, x_m$ produces the output $y$.
    At the same time, an auditor $\mathcal{P}^A$ will reject $y$ (except with negligible probability), if $\tau$ is not a valid transcript of an evaluation of $\mathcal{C}$ or if $\mathcal{C}\left(x_1, \ldots, x_m\right) \neq y$.
    \end{definition}

    As an alternative definition for publicly verifiable \ac{MPC} we observe the definition of~\cite{ozdemirExperimentingCollaborativeZkSNARKs2022} for publicly auditable \ac{MPC} from collaborative proofs, which can be seen as a sort of unification between \ac{VC} and \ac{MPC}:
    \begin{definition}[Publicly auditable \ac{MPC} from collaborative proofs for $t$(-out-of-$N$) malicious parties~\cite{ozdemirExperimentingCollaborativeZkSNARKs2022}]\label{def:pa-mpc-ozdemir}
    Given a commitment scheme \commit, we define a publicly auditable \ac{MPC} for a function $f: X^N \rightarrow Y$ as a 3-tuple of \ac{ppt} algorithms:
    \begin{itemize}
        \item $\setup(f, 1^\securityParameter) \rightarrow \publicParameters$: given a function $f$ and security parameter \securityParameter, this returns public parameters \publicParameters;
        \item $\Pi(\publicParameters, \vec{x}) \rightarrow (y, \zkproof)$: Given the public parameters \publicParameters, and input $\vec{x}$, all parties execute a protocol to compute the function output $y$ and a proof \zkproof attesting to the correctness thereof;
        \item $\verify(\publicParameters, \vec{c}, y, \zkproof) \rightarrow \{0,1\}$: Given the public parameters \publicParameters, input commitments $\vec{c}$, output value $y$ and proof \zkproof, returns 1 if $f(\vec{x}) = y$, and 0 otherwise.
    \end{itemize}
    Such a scheme is secure against $t$ malicious parties if:
    \begin{itemize}
        \item $\Pi$ (when only considering its output $y$) is a secure-with-abort \ac{MPC} for evaluating $f$, against $t$ malicious parties; and
        \item The above collection of algorithms (when only considering the $\zkproof$ output value of $\Pi$) is a collaborative \ac{zk-SNARK} for the relation $\mathcal{R}_f = \left\{(\vec{c}, y; \vec{x}, \vec{r}: f(\vec{x}) = y \land c_i = \commit(x_i, r_i), \forall i)\right\}$, secure against $t$ malicious provers.
    \end{itemize}
    \end{definition}

    \subsection{Prior definitions for verifiable HE}\label{subsec:formal-he}
    As observed in~\cite{viandVerifiableFullyHomomorphic2023}, there is a wide variety of definitions for verifiable \ac{HE}.

    Some works, e.g.,~\cite{liPrivacyPreservingHomomorphicMACs2018}, rely on typical \ac{MAC} unforgeability definitions, likely due their heavy reliance on homomorphic \acp{MAC} to construct a verifiable \ac{HE} scheme.
    Generally speaking, the defined homomorphic \acp{MAC} should satisfy: \emph{authentication correctness}, \emph{evaluation correctness}, \emph{efficiency}/\emph{succinctness}, and \emph{unforgeability}.
    We refer to~\cite{liPrivacyPreservingHomomorphicMACs2018} for the exact, formal definitions hereof.
    In the case of \textit{Encrypt-and-\acs*{MAC}} constructions \emph{privacy} should also be guaranteed.
    For other constructions, privacy is already guaranteed by the \ac{HE} scheme.
    The downside of these definitions is that they focus purely on the security of the homomorphic \ac{MAC} and do not describe the properties of the complete verifiable \ac{HE} system.

    Alternatively, other works, e.g.,~\cite{fioreEfficientlyVerifiableComputation2014}, adapt the \ac{VC} definitions of~\cite{gennaroNoninteractiveVerifiableComputing2010}, towards \ac{HE}, i.e., their schemes satisfy \emph{correctness}, \emph{security}, \emph{privacy}, and \emph{efficiency} definitions.
    While this does describe the properties of the system as a whole, it leaves out some crucial properties present in modern \ac{HE} schemes, such as approximate evaluation.

    A set of generic definitions for verifiable \ac{HE} schemes is, to the best of our knowledge, first defined in~\cite{viandVerifiableFullyHomomorphic2023} (see \cref{def:he}).

    We observe that such a scheme is only publicly verifiable if the secret key is not needed in order to execute $\verify$.
    Regarding the other properties, we observe that the authors split the correctness notion as seen before, into \emph{correctness} and \emph{completeness} (\cref{def:he-correctness,def:he-completeness}).
    Correctness guarantees that the decrypted, evaluated result is exactly equal to the evaluation of $f(x)$, whereas completeness only verifies that that the encrypted result is indeed an evaluation of $f$ on the encrypted version of $x$.
    For approximate \ac{HE} schemes, correctness cannot be guaranteed exactly, and another definition of approximate correctness is required~\cite{viandVerifiableFullyHomomorphic2023}.
    Next to this, a verifiable \ac{HE} scheme should be sound and provide privacy.
    Soundness (\cref{def:he-soundness}) is defined as the equivalent of security in \ac{VC} and privacy (it is referred to as security in~\cite{viandVerifiableFullyHomomorphic2023}) can be extended from the definition of \textsf{IND-CCA1} (\cref{def:he-privacy}).

    Finally, we note that~\cite{viandVerifiableFullyHomomorphic2023} also gives a number of additional definitions for properties that may be useful in specific settings, such as server privacy (for additional inputs provided by the server) and circuit privacy (or function privacy).
    \begin{definition}[Verifiable \ac{HE}~\cite{viandVerifiableFullyHomomorphic2023}]\label{def:he}
    A verifiable \ac{HE} scheme is a 5-tuple of \ac{ppt} algorithms:
    \begin{itemize}
        \item $\keygen(f, 1^\securityParameter) \rightarrow (\publicKey, \secretKey)$: given a function $f$ and security parameter \securityParameter, outputs a public key \publicKey and secret key \secretKey;
        \item $\encrypt(\publicKey/\secretKey, x) \rightarrow (c_x, \tau_x)$: given either the public key \publicKey or secret key \secretKey and an input $x$, returns a ciphertext $c_x$ and corresponding tag $\tau_x$;
        \item $\eval(\publicKey, c_x) \rightarrow (c_y, \tau_y)$: given the public key \publicKey, and input ciphertext $c_x$, returns output ciphertext $c_y$ and corresponding tag $\tau_y$;
        \item $\verify(\secretKey, c_y, \tau_x, \tau_y) \rightarrow \{0,1\}$: given the secret key \secretKey, output ciphertext $c_y$, and tags $\tau_x$ and $\tau_y$, returns 1 if $c_y$ is a valid output ciphertext, and 0 otherwise;
        \item $\decrypt(\secretKey, c_y) \rightarrow y$: given the secret key \secretKey and output ciphertext $c_y$, returns the decrypted output $y$.
    \end{itemize}
    \end{definition}

    \begin{definition}[Correctness~\cite{viandVerifiableFullyHomomorphic2023}]\label{def:he-correctness}
    A verifiable \ac{HE} scheme is correct if for all functions $f$, with $\keygen(f, 1^\securityParameter) \rightarrow (\publicKey, \secretKey)$, \\ such that $\forall x \in Domain(f)$, if $\encrypt(\publicKey/\secretKey, x) \rightarrow (c_x, \tau_x)$, \\ $\eval(\publicKey, c_x) \rightarrow (c_y, \tau_y)$ then $\Pr[\decrypt(\secretKey, c_y) = f(x)] = 1$.
    \end{definition}

    \begin{definition}[Completeness~\cite{viandVerifiableFullyHomomorphic2023}]\label{def:he-completeness}
    A verifiable \ac{HE} scheme is complete if for all functions $f$, with $\keygen(f, 1^\securityParameter) \rightarrow (\publicKey, \secretKey)$, such that $\forall x \in Domain(f)$, if $\encrypt(\publicKey/\secretKey, x) \rightarrow (c_x, \tau_x)$, \\ $\eval(\publicKey, c_x) \rightarrow (c_y, \tau_y)$ then $\Pr[\verify(\secretKey, c_y, \tau_x, \tau_y) = 1] = 1$.
    \end{definition}

    \begin{definition}[Soundness~\cite{viandVerifiableFullyHomomorphic2023}]\label{def:he-soundness}
    A verifiable \ac{HE} scheme is sound, if for any \ac{ppt} adversary $\mathcal{A} = (\mathcal{A}_1, \mathcal{A}_2)$ such that for all functions $f$, if $\keygen(f, 1^\securityParameter) \rightarrow (\publicKey, \secretKey)$, $\mathcal{A}_1^{\mathcal{O}_{\encrypt},\mathcal{O}_{\decrypt}}(\publicKey) \rightarrow x$, \\ $\encrypt(\publicKey/\secretKey, x) \rightarrow (c_x, \tau_x)$, and $\mathcal{A}_2^{\mathcal{O}_{\encrypt},\mathcal{O}_{\decrypt}}(c_x, \tau_x) \rightarrow (c_y, \tau_y)$ \\ then $\Pr[\verify(\secretKey, c_y, \tau_x, \tau_y) = 1 \land \decrypt(\secretKey, c_y) \neq f(x)] \leq \textsf{negl}(\securityParameter)$.
    \end{definition}

    \begin{definition}[Privacy~\cite{viandVerifiableFullyHomomorphic2023}]\label{def:he-privacy}
    A verifiable \ac{HE} scheme satisfies the privacy property, if for any \ac{ppt} adversary $\mathcal{A} = (\mathcal{A}_1, \mathcal{A}_2)$ such that for all functions $f$, if $\keygen(f, 1^\securityParameter) \rightarrow (\publicKey, \secretKey)$, \\ $\mathcal{A}_1^{\mathcal{O}_{\encrypt},\mathcal{O}_{\decrypt}}(\publicKey, f, 1^\securityParameter) \rightarrow (x_0, x_1)$, \\ $\{0,1\} \xrightarrow{R} b$, $\encrypt(\publicKey/\secretKey, x_b) \rightarrow (c_b, \tau_b)$ \\ then $2|  \Pr[\mathcal{A}_2^{\mathcal{O}_{\encrypt}}(c_x) = b] - \frac{1}{2}| \leq \textsf{negl}(\securityParameter)$.
    \end{definition}

    In \cref{def:he-soundness,def:he-privacy}, we assume that the adversary has access to a decryption oracle $\mathcal{O}_\text{dec}$ and an encryption oracle $\mathcal{O}_\text{enc}$ where appropriate.
    The decryption oracle $\mathcal{O}_\text{dec}(c_y, \tau_x, \tau_y)$ returns $\bot$ if $\verify(\secretKey, c_y, \tau_x, \tau_y) = 0$, and $\decrypt_{\secretKey}(c_y)$ otherwise.
    Similarly, the encryption oracle $\mathcal{O}_\text{enc}(x)$ returns $\encrypt(\publicKey/\secretKey, x)$ for any valid input $x$.

\end{document}